\documentclass[a4paper,11pt]{article}
\pdfoutput=1 

\usepackage{jheppub} 

\usepackage[T1]{fontenc} 

\def\N{{\mathcal N}}

\def\F{{\mathcal F}}

\def\PW{\text{Pilch-Warner }}

\newcommand{\dd}{\mbox{d}}
\newcommand{\diag}{\mbox{diag}}
\def\axs{AdS_5\times S^5}

\title{\boldmath Holographic mesons in global Pilch-Warner background geometry}

\author[a,b]{R. C. Rashkov}
\author[b]{and T. Vetsov}

\affiliation[a]{Institute for Theoretical Physics, Vienna University of Technology, Wiedner Hauptstr. 8-10, 1040 Vienna, Austria}
\affiliation[b]{Department of Physics, Sofia University, 5 James Bourchier Blvd., BG-1164 Sofia, Bulgaria}

\emailAdd{rash@hep.itp.tuwien.ac.at}
\emailAdd{vetsov@phys.uni-sofia.bg }

\abstract{In this paper we study D-brane scalar fluctuations in global Pilch-Warner background geometry. We consider configurations of the probe branes compatible with the kappa symmetry preserving condition and the brane classical equations of motion. The corresponding meson spectra, obtained by the fluctuations along the transverse brane directions, admit equidistant structure for the higher modes, but some of them show additional shifts in their ground states.}


\begin{document}
\maketitle
\flushbottom

\section{Introduction}
\label{sec: Introduction}
The AdS/CFT correspondence is a fascinating duality relating 10 dimensional IIB string theory in the weak coupling regime to a four dimensional $SU(N)$ gauge field theory with strong coupling constant, and vice-versa. In this case the gauge field theory lives on the boundary of the spacetime where the strings move. This correspondence gives us the opportunity to study non-perturbative phenomena in Yang-Mills theory with tools available in the classical superstring theory and supergravity.
\\
\indent In the original Maldacena setup \cite{Maldacena:1998} there is an $\mathcal{N}=4$ supersymmetric Yang-Mills theory on the gauge side of the correspondence, and a stack of parallel $N_c$ D3-branes on the string side of the correspondence. Here both endpoints of the strings are attached to the same stack of D3-branes, which allows only adjoint fields. Adding flavours or fundamental matter like quarks can be achieved by introducing a separate stack of $N_f$ D7 probe branes. Such configurations result in appearance  of $SU(N_f)$ global flavor symmetry \cite{Karch:2002}. Now the endpoints of the fundamental strings lie on a different stack of branes. Since they are separated in some directions at finite length, and therefore the strings have finite energy, the quark mass is given by the separation distance times the string tension: $m_q=L/2 \pi \alpha'$. If we consider the case where $N_f\ll N_c$ and then take large number of D3-branes, we end up with a strongly coupled dual gauge theory, and a stack of D3-branes sourcing the background geometry. In the limit of large $N_c$ the stack of D3-branes can effectively be replaced by the $AdS_5\times S^5$ space. This setup allows us to study the $N_f$ D7-branes in the probe limit, where the energy density of the stack of D7-branes does not backreact on the background geometry.
\\
\indent Variety of supersymmetric meson spectra were found in \cite{Myers1:2003, FilevGlobal:2010} (for an extensive review see \cite{ErdmengerReview:2008}. In order to find realistic string theory description of QCD and the Standard model we have to reduce the supersymmetry to some level, but still keeping control on the theory. A way to achieve this goal is to consider deformations of the initial $AdS_5\times S^5$ geometry \cite{Johnson2:2011, Filev:2013}, or involvement of external magnetic or electric fields \cite {Filev:2008, Filev2:2010, Filev:2009, FilevDis:2008}. Such configurations will break the supersymmetry and theories with less supersymmetry will emerge. In this context \PW geometry \cite{PW1:2000, PW2:2002, Rash:2003} is a fine example of such deformed geometry. It is a solution of five-dimensional $\mathcal{N} = 8$ gauged supergravity lifted to ten dimensions and preserving 1/4 of the original supersymmetry in its infrared critical point.
\\
\indent The rich structure of the theory on both sides of the AdS/CFT correspondence promises interesting results. The recent revival interest to the holographic dual of Pilch-Warner geometry \cite{Zarembo:2015, Zarembo:2014, Zarembo2:2014, Buchel1:2013, Buchel2:2013} also motivates our study.
\\
\indent This paper is organized as follow. First we give some
details about Pilch-Warner background. After that in section \ref{sec: Kappa symmetry matrix} we find the explicit form of the kappa symmetry matrix for both the D5 and D7 probe branes. Then we solve the kappa symmetry preserving condition. Its solution allows us to find embeddings of the branes compatible with the kappa symmetry and the classical equations of motion. We also prove that the kappa symmetry preserves exactly 1/2 of the spinor degrees of freedom for both D5 and D7.
\\
\indent In section \ref{sec:D7-Scalar-Fluctuations} we consider the kappa symmetric D7-brane embedding and study the corresponding spectrum of scalar fluctuations. We show that the ground state of
the spectra along the $\phi$ and $\theta$ directions are equidistant for the higher modes. However, an unexpected shift appears in the ground state of the spectrum along the $\theta$ direction.
\\
\indent Analogously, in section \ref{sec:D5-Scalar-Fluctuations} we study the kappa symmetric D5 embedding and analyze the spectrum of the corresponding fluctuations.
\\
\indent We conclude by briefly discussing the results in section \ref{sec:Conclusion}.
\\
\indent Finally, there are number of appendices. Appendix \ref{secA:Appendix-A} contains the explicit form of the R-R and NS-NS potentials written in global \PW coordinates. In Appendices \ref{secB:Appendix-B} and \ref{secC:Appendix-C} we give the explicit representation of the ten dimensional Dirac gamma matrices and the vielbein coefficients of the PW metric. In Appendices \ref{sec: Appendix-D} and \ref{secC:Appendix-E} we present in details the form of the kappa symmetry matrices for D7 and D5 probe branes.
\section{General Setup}
\label{sec: General Setup}
This section is intended to briefly introduce the the Pilch-Warner background geometry, D-branes embeddings and their field content. First we introduce some details about the ten dimensional PW background. Then we introduce D5 and D7 probes. We briefly discus the form of the non-trivial Ramond-Ramond (R-R) and Neveu-Schwarz (NS-NS) fields entering the D-branes effective actions.
\subsection{\PW geometry}
\label{subsec: PW Geometry}
Pilch-Warner geometry is a solution of five-dimensional $\N = 8$ gauged supergravity lifted to ten dimensions. In the UV critical point it gives the maximally supersymmetric
$\axs$, while in the infrared point (IR) of the flow it describes warped $AdS_5$ times
squashed $S^5$. In this paper we will restrict ourself to the IR critical point. An important feature on the gravity side is that it preserves 1/8 supersymmetry everywhere along the flow, while at IR fixed point it is enhanced to 1/4.
On the SYM side the IR point corresponds to the large $N$ limit of the
superconformal $\N = 1$ theory of Leigh-Strassler \cite{Strassler:1995}.
\\
\indent The warped $AdS_5$ part of the ten-dimensional PW metric is written by
\begin{equation}
\dd s_{1,4}^2 = {\Omega ^2}\,\left({e^{2A}}\,{\rm{d}}{s^2}({\mathbb{M}^4}) + {\rm{d}}{r^2}\right)\,,
\end{equation}
while the squashed five-sphere part is
\begin{align}
ds_5^2 &= \frac{{L_0^2\,{\Omega ^2}}}{{{{\bar \rho }^{{\kern 1pt} 2}}\,{{\cosh }^2}\chi }}\,\left[ {d{\theta ^2} + \frac{{{{\bar \rho }^{\,6}}\,{{\cos }^2}\theta }}{X}\,\left( {\sigma _1^2 + \sigma _2^2} \right) + \frac{{{{\bar \rho }^{12}}\,{{\sin }^2}2\theta }}{{4\,{X^2}}}} \right.\,{\left( {{\sigma _3} + \frac{{2 - {{\bar \rho }^6}}}{{2\,{{\bar \rho }^6}}}\,d\phi } \right)^2}\nonumber\\
&\left. { + \frac{{{{\bar \rho }^{{\kern 1pt} 6}}\,{{\cosh }^2}\chi }}{{16\,{X^2}}}\,{{(3 - \cos 2\theta )}^2}\,{{\left( {d\phi  - \frac{{4\,{{\cos }^2}\theta }}{{3 - \cos 2\theta }}\,{\sigma _3}} \right)}^2}} \right]\,.
\end{align}
Here the function $\Omega$ is called the warp factor:
\begin{equation}
{\Omega ^2} = \frac{{{X^{1/2}}\,\cosh \chi }}{{\bar \rho }}\,,\quad
 X(r,\theta ) = {\cos ^2}\theta + {\bar \rho ^{\,6}}\,{\sin ^2}\theta \,.
\end{equation}
Our left-invariant one-forms ${\sigma }_{i}$ satisfy $\text{d}{{\sigma }_{i}}={{\varepsilon }_{ijk}}\,{{\sigma }_{j}}\wedge {{\sigma }_{k}}$, so that $\text{d}\tilde{\Omega} _{3}^{2}={{\sigma }_{i}}\,{{\sigma }_{i}}$ is the metric on the unit 3-sphere. We can choose them as
\begin{subequations}
\begin{align}
{\sigma _1} &= \frac{1}{2}\,(\sin {\beta}\,{\rm{d}}{\alpha} - \cos {\beta}\,{\rm{sin}}{\alpha}\,{\rm{d}}{\gamma})\,,
\\
{\sigma_2} &= - \frac{1}{2}\,(\cos {\beta}\,{\rm{d}}{\alpha} + \sin {\beta}\,{\rm{sin}}{\alpha}\,{\rm{d}}{\gamma})\,,
\\
{\sigma_3} &= \frac{1}{2}\,({\rm{d}}{\beta} + \cos {\alpha}\,{\rm{d}}{\gamma}) \,.
\end{align}
\end{subequations}
At the IR point one has $r\to -\infty $, $\chi = \operatorname{arccosh} (2/\sqrt{3})$, $\bar{\rho }={{2}^{1/6}}$, and $A(r)=r/L$, thus the metric can be written in the form
\begin{equation}
\label{IR Warped AdS PW Metric}
\dd s_{1,4}^2 = {\Omega ^2}\,\left({e^{\frac{{2\,r}}{L}}}\,\dd {s^2}({\mathbb{M}^4}) + \dd{r^2}\right)\,,
\end{equation}
\begin{align}
\label{IR Squashed Sphere PW Metric}
{\rm{d}}s_{5}^2 &= \frac{2}{3}\,{L^2}\,{\Omega ^2}\, \left[ {{\rm{d}}{\theta ^2} + \frac{{4\,{{\cos }^2}\theta }}{{3 - \cos 2\theta }}\left( {\sigma _1^2 + \sigma _2^2} \right) + \frac{{4\,{{\sin }^2}2\theta }}{{{{(3 - \cos 2\theta )}^2}}}\,\sigma _3^2 + }
{\frac{2}{3}\,{{\left( {{\rm{d}}\phi \, - \frac{{4\,{{\cos }^2}\theta }}{{3 - \cos 2\theta }}\,{\sigma _3}} \right)}^2}} \right]\,.
\end{align}
where the $AdS$ radius $L$ is given in terms of the $AdS$ radius $L_0$ of the UV spacetime by $L=(3/{{2}^{5/3}})\,{{L}_{0}}$. As shown in \cite{Johnson1:2002} there is a natural global $U{{\left( 1 \right)}_{{\beta}}}$ action  ${\beta}\to {\beta}+const$, which rotates ${{\sigma }_{1}}$ into ${{\sigma }_{2}}$, but leaves ${{\sigma }_{3}}$
invariant. We adopt the set up where the $S^3$ Euler angle ${\beta}\to {\beta}+2\,\phi$ is shifted to give a solution with a global $U(1)_R = U(1)_\phi$ symmetry. Performing this coordinate transformation on the solution (\ref{IR Warped AdS PW Metric}) and (\ref{IR Squashed Sphere PW Metric}) we arrive at the final result for the \PW metric in global coordinates at the IR fixed point:
\begin{equation}
{\rm{d}}s_{1,4}^2(IR) = {L^2}\,{\Omega ^2}\,\left( { - {{\cosh }^2}\rho \,{\rm{d}}{\tau ^2} + {\rm{d}}{\rho ^2} + {{\sinh }^2}\rho \,\rm{d}\Omega^2_3} \right)\,,
\end{equation}
\begin{align}
{\rm{d}}s_5^2(IR) &= \frac{2}{3}\,{L^2}\,{\Omega ^2}\,\left[ {{\rm{d}}{\theta ^2} + \frac{{4\,{{\cos }^2}\theta }}{{3 - \cos 2\theta }}\left( {\sigma _1^2 + \sigma _2^2} \right) + \frac{{4\, {{\sin }^2}2\theta }}{{{{(3 - \cos 2\theta )}^2}}}\,{{({\sigma _3} + {\rm{d}}\phi )}^2}} \right.\nonumber\\
\nonumber\\
&\left. { + \frac{2}{3}\,{{\left( {\frac{{1 - 3\,\cos 2\theta }}{{\cos 2\theta  - 3}}} \right)}^2}\,{{\left( {{\rm{d}}\phi  - \frac{{4\,{{\cos }^2}\theta }}{{1 - 3\,\cos 2\theta }}\,{\sigma _3}} \right)}^2}}  \right]\,,
\end{align}
where $\dd\Omega _3^2 = \dd\phi _1^2 + {\sin ^2}{\phi _1}\,\left(\dd\phi _2^2 + {\sin ^2}{\phi _2}\,\dd\phi _3^2\right)$ is the metric on the unit 3-sphere, and
\begin{equation}\label{The warp factor at the IR point}
  {\Omega ^2} = \frac{{{2^{1/3}}}}{{\sqrt 3 }}\,\sqrt {3 - \cos 2\theta }
\end{equation}
represents the warp factor at the IR point.
\\
\indent As a supergravity solution the \PW background includes non-trivial Ramond-Ramond (R-R) and Neveu-Schwarz (NS-NS) form fields. The field content of the IIB string theory contains the following R-R potentials: $C_0,\,C_2,\,C_4,\,C_6,\,C_8$ with their corresponding field strengths, and the NS-NS Kalb-Ramond two-form $B_2$. These fields satisfy certain Bianchi identities and equations of motion listed below: \cite{Mateos:2010}:
\begin{subequations}\label{Bianchi-Identity-For-The-R-R-Forms}
\begin{equation}
\label{C0-Form-Definition}
\Phi=C_0=0,\,\, F_1=dC_0=0\,,
\end{equation}
\begin{equation}
C_2=\Re e(A_2),\,\, B_2=\Im m(A_2)\,,
\end{equation}
\begin{equation}
H_3=dB_2,\,\, F_3=dC_2-C_0\wedge H_3=dC_2\,,
\end{equation}
\begin{equation}
\rm{d}\textit{F}_3 = \rm{d}\textit{H}_3 = 0,\,\, \rm{d}\textit{F}_5 = \textit{H}_3 \wedge \textit{F}_3\,,
\end{equation}
\begin{equation}
\rm{d}( \star \textit{F}_3) =  - \textit{H}_3 \wedge \textit{F}_5,\,\, \rm{d}( \star \textit{H}_3) = \textit{F}_3 \wedge \textit{F}_5,\,\,\textit{F}_5 =  \star \textit{F}_5\,,
\end{equation}
\begin{equation}
dC_4+d\tilde{C}_4=F_5+C_2\wedge H_3\,,
\end{equation}
\begin{equation}
F_7=\star F_3=dC_6-C_4\wedge H_3\,,
\end{equation}
\begin{equation}
\label{C8-Form-Definition}
F_9=\star F_1=0=dC_8-C_6\wedge H_3=C_6\wedge H_3,\,\,\chi=C_8=0\,.
\end{equation}
\end{subequations}
Here the field strengths are defined in terms of the corresponding potentials as
\begin{equation}
    \textit{H}_3 \equiv \rm{d}\textit{B}_2,\quad \textit{F}_p \equiv \rm{d}{\textit{C}_{p - 1}} - {\textit{C}_{p - 3}} \wedge {{\rm\textit{H}}_3}.
\end{equation}
In this setup the axion/dilaton system of scalars (\ref{C0-Form-Definition}) and (\ref{C8-Form-Definition}) is trivial along the flow. We also have an ansatz for the self-dual five form
\begin{equation}
    {F_5} =  - \frac{{{2^{5/3}}}}{3}\,{L^4}\,\cosh \rho \,{\sinh ^3}\rho \,(1 +  \star )\,{\rm{d}}\tau  \wedge {\rm{d}}\rho  \wedge \epsilon(S_\phi ^3)\,,
\end{equation}
where $ \epsilon(S_\phi ^3)= \sin^2\phi_1\,\sin\phi_2\,\rm{d}\phi_1\wedge\rm{d}\phi_2\wedge\rm{d}\phi_3$ is the volume element of the unit 3-sphere $S_\phi ^3$, and $\star$ represents the Hodge star operator. The ansatz for the 2-form potential $A_2$ at the IR point is given by
\begin{equation}
    {A_2}(IR) =  C_2+i\,B_2=- \frac{i}{2}\,{e^{ - 2\,i\,\phi }}\,L_0^2\,\cos \theta \left( {{\rm{d}}\theta  - \frac{{2\,i\,\sin 2\theta }}{{3 - \cos 2\theta }}{\,(\sigma _3+\rm{d}\phi)}} \right) \wedge ({\sigma _1} + i\,{\sigma _2})\,.
\end{equation}
Two additional constraints are necessary for (\ref{C0-Form-Definition}) and (\ref{C8-Form-Definition}) to be consistent, namely
\begin{equation}
    \textit{F}_3 \wedge  \star \textit{H}_3 = 0\quad \text{and}\quad \textit{F}_3 \wedge \star \,\textit{F}_3 = \textit{H}_3 \wedge  \star {{\rm\textit{H}}_3}\,.
\end{equation}
The explicit form of the RR and the NS-NS potentials in PW satisfying the equations above can be found in Appendix \ref{secA:Appendix-A}.
\subsection{D-brane embeddings and their field content}
As we mentioned in the previous subsection the \PW background includes non-trivial Ramond-Ramond (R-R) and Neveu-Schwarz (NS-NS) form fields. Their pullbacks on the D-brane world volume will induce certain terms in the action. The full D-brane effective action, governing the low energy dynamics of the D$p$-branes, is given by
\begin{equation}\label{The full D brane effective action}
{S_{{D_p}}} =  - {T_p}\,\int\limits_{} {{d^{p + 1}}\xi \,{e^{ - \Phi }}\,\sqrt { - \det \left( {{\cal G} + {\cal F}} \right)} }  + {\mu _p}\,\int\limits_{} {\sum\limits_n^{} {P[{{ C}_{(n)}}] \wedge {e^{\cal F}}} }\,,
\end{equation}
where ${\cal F} = {{\cal B}_{(2)}} + 2\,\pi \,\alpha '\,{F_{(2)}}$ is the invariant gauge two-form, $F_{(2)}$ is the world volume gauge field, and ${\cal{B}}_{(2)}$ is the pullback of the Kalb-Ramond field. The relation between the D$p$-brane tension $T_p$\footnote{First calculated in \cite{Polchinski:1995}.} and its charge $\mu_p$ is fixed by the supersymmetry: $\mu_p=\pm T_p$. We also have the dilaton $\Phi$ and the pullback ${\cal G}$ of the background metric:
\begin{equation}
   \mathcal{G}_{ab}
    =G_{AB}\,\frac{\partial X^A}{\partial \xi^a}\,
    \frac{\partial X^B}{\partial \xi^b}\,.
\end{equation}
The indices $a,\,b=0,\dots,p$ span the world volume of the D$p$-brane, while $A,\,B=0,\dots,9$ span the whole spacetime. Often it is more useful to work in static gauge, $X^a=\xi^a$, where the pullback is given by
\begin{equation}\label{Pullback of the PW metric in static gauge}
\mathcal{G}_{ab}
=g_{ab}+G_{mn}\,
\frac{\partial X^m}{\partial \xi^a}\,\frac{\partial X^n}{\partial \xi^b}\,.
\end{equation}
Here $g_{ab}$ is the induced metric on the D-brane world volume and $G_{mn}$ are the metric components in front of the transverse coordinates governing the D$p$-brane fluctuations.
\\
\indent The first term in (\ref{The full D brane effective action}) is the Dirac-Born-Infeld (DBI) action, and the second term is the Wess-Zumino (WZ) action \cite{Mateos:2010, Polchinski:1995}. The later has to be expanded in formal series where only $(p+1)$-forms are selected. For example the WZ part of the action for the D7-brane takes the form
\begin{equation}
S_{WZ}= - {T_7}\,\int \left(P[C_8]-P[C_6]\wedge \F+\frac{1}{2}\,P[C_4]\wedge\F\wedge\F+\cdots\right)\,,
\end{equation}
where $P$ denotes the pullback of the bulk spacetime tensor to the world-volume of the brane. In general one has
\begin{equation}\label{Pullback of the RR potential}
P{[{{C}_{p + 1}}]_{{a_1} \ldots {a_{p + 1}}}} = \frac{1}{{(p + 1)!}}\,{\varepsilon ^{{a_1} \ldots {a_{p + 1}}}}\,{\partial _{{a_1}}}\,{X^{{A _1}}} \ldots {\partial _{{a_{p + 1}}}}{X^{{A _{p + 1}}}}\,{C_{{A _1} \ldots {A _{p + 1}}}}(X)\,,
\end{equation}
where ${\varepsilon ^{{a_1} \ldots {a_{p + 1}}}}$ is the antisymmetric Levi-Chevita symbol.
\\
\indent In what follows we would like our D7 probe to extend on the whole warped $AdS_5$ part of the space, while wrapping a three sphere of the squashed $S^5$ spanned by the angles ($\alpha, \beta,\gamma$). This leave us with the following general D7-brane embedding in global \PW coordinates:
\begin{equation}
{\xi ^a} = (\tau ,\rho ,{\phi _1},{\phi _2},{\phi _3},\alpha ,\beta ,\gamma ),\quad \theta  = \theta ({\xi ^a}),\quad \phi  = \phi ({\xi ^a})\,,
\end{equation}
where $\xi^a$ are the world volume coordinates, and ($\theta$, $\phi$) are the normal coordinates of the brane.
\\
\indent The most general embedding of the D5-brane, that we are going to consider, is given by
\begin{align}
{\xi ^a} = (\tau ,\rho ,{\phi _2},{\phi _3},\beta ,\gamma ),\quad {\phi _1} = {\phi _1}({\xi ^a}),\quad \theta  = \theta ({\xi ^a}),\quad \alpha  = \alpha ({\xi ^a}),\quad \phi  = \phi ({\xi ^a})\,,
\end{align}
where $\xi^a$, $a=0,\dots,6$, are the D5 world volume coordinates. In global \PW this corresponds to a D5-brane embedding wrapping warped $AdS_4$ subspace and a two-sphere of the squashed $S^5$.
\section{Kappa symmetry matrix}
\label{sec: Kappa symmetry matrix}
In this section we consider the implications and the importance of the local fermionic kappa symmetry. Our goal is to find kappa symmetric embeddings of the $D5$ and $D7$ probe branes by solving the kappa symmetry preserving condition.
\\
\indent Our D-brane effective action has an additional local fermionic gauge symmetry -- kappa symmetry. It implies that half of the components of the background killing spinor $\epsilon$ are actually gauge degrees of freedom and can be gauged away. This symmetry is required for the preservation of spacetime covariance and supersymmetry. Kappa symmetry invariance is achieved whenever the background is an on-shell supergravity background, which is the case of the PW geometry.  The general formalism to study supersymmetric bosonic world volume solitons requires that any such configuration must satisfy the kappa symmetry preserving condition \cite{Bergshoeff:1997}
\begin{equation}\label{kappa symmetry preserving condition}
  \Gamma_k\,\varepsilon=\varepsilon \,,
\end{equation}
where $\Gamma_\kappa^2=1$ and $\varepsilon$ is a Weyl spinor\footnote{If we require supersymmetry invariance the spinor $\varepsilon$ becomes the background Killing spinor $\epsilon$.}. It is important to stress that kappa symmetry invariance is necessary to define a supersymmetric field theory on the brane, but not sufficient. To determine the supersymmetric configurations of the probe brane the kappa projection condition (\ref{kappa symmetry preserving condition}) must be compatible with some additional projection conditions for the supergravity background Killing spinor.
\\
\indent Following \cite{Simon:2011} we have the following expression for the $(p+1)$-form kappa symmetry matrix:
\begin{equation}
  {({\Gamma _\kappa })_{(p + 1)}} = \frac{1}{{\sqrt { - \det \left( {{\cal G} + {\cal F}} \right)} }}\,\sum\limits_{\ell  = 0}^{k + 1} {{\gamma _{(2\,\ell )}}\,\sigma _3^\ell \, \wedge {e^{\cal F}}\,i\,{\sigma _2}} ,\qquad p = 2\,k + 1,
\end{equation}
where
\begin{equation}
  {\cal F} = 2\,\pi \,\alpha '\,{F_{(2)}} + {{\cal B}_{(2)}}\,,\quad {{\cal B}_{(2)}} = P[{B_{(2)}}]\,,\quad {{\cal G}} = P[{G}]\,.
\end{equation}
The notation ${\gamma _{(\ell )}}$ stands for the wedge product of the following one-forms:
\begin{equation}
  {\gamma _{(1)}} \equiv d{\xi ^a}\,{\gamma _a} = d{\xi ^a}\,{\partial _a}{X^M}\,E_M^I(X)\,{\Gamma _I}\,.
\end{equation}
We can expand $e^{\cal F}$ in a formal series of the form
\begin{equation}
  {e^{\cal F}} = 1 + {\cal F} + \frac{1}{{2!}}\,{\cal F} \wedge {\cal F} + \frac{1}{{3!}}\,{\cal F} \wedge {\cal F} \wedge {\cal F} + \frac{1}{{4!}}\,{\cal F} \wedge {\cal F} \wedge {\cal F} \wedge {\cal F} +  \cdots
\end{equation}
Therefore the kappa matrix takes the form
\begin{equation}\label{General form of the Kappa symmetry matrix}
{({\Gamma _\kappa })_{(p + 1)}} = \frac{1}{{\sqrt { - \det \left( {{\cal G} + {\cal F}} \right)} }}\,\sum\limits_{\ell  = 0}^{k + 1} {{\gamma _{(2{\kern 1pt} \ell )}}\,\sigma _3^\ell \, \wedge \left( {1 - {\cal F} + \frac{1}{{2!}}\,{{\cal F}^2} - \frac{1}{{3!}}\,{{\cal F}^3} + \frac{1}{{4!}}\,{{\cal F}^4} +  \cdots } \right) \,i\,{\sigma _2}} \,,
\end{equation}
where ${{\cal F}^2} = {\cal F} \wedge {\cal F}$.
\\
\indent The calculation of the Killing spinor in global PW geometry and the analysis of the supersymmetric brane embeddings will be left for a future work. In what follows we consider only kappa symmetric embeddings compatible with the D5- and D7-brane equations of motion.
\subsection{Kappa summetry for the D7-brane}
For the D7-brane one has $p=8, k=3$, and only 8-form terms are selected. Thus (\ref{General form of the Kappa symmetry matrix}) takes the form
\begin{equation}\label{D7 Kappa Matrix 8-Form}
{({\Gamma _\kappa })_{(8)}} = \frac{1}{{\sqrt { - \det \left( {{\cal G} + {\cal F}} \right)} }}\,\left( {{\gamma _{(8)}}\,\sigma _3^4 + {\gamma _{(6)}}\,\sigma _3^3 \wedge {\cal F} + \frac{{{\gamma _{(4)}}\,\sigma _3^2}}{{2!}} \wedge {{\cal F}^2} + \frac{{{\gamma _{(2)}}\,\sigma _3^1}}{{3!}} \wedge {{\cal F}^3} +  \cdots } \right)\,i\,{\sigma _2}\,.
\end{equation}
In order to explicitly calculate the D7 kappa symmetry matrix we begin with the most general D7 embedding in terms of global \PW coordinates:
\begin{equation}
{\xi ^a} = (\tau ,\rho ,{\phi _1},{\phi _2},{\phi _3},\alpha ,\beta ,\gamma ),\quad \theta  = \theta ({\xi ^a}),\quad \phi  = \phi ({\xi ^a})\,.
\end{equation}
The analysis will greatly simplify if we set the world volume gauge field $F=0$. The wedge products of the $B_{(2)}$-form pullback terminates at third order $\mathcal{B}_{(2)}\wedge\mathcal{B}_{(2)}\wedge\mathcal{B}_{(2)}=0$, and kappa symmetry 8-form matrix (\ref{D7 Kappa Matrix 8-Form}) becomes
\begin{equation}
{({\Gamma _\kappa })_{(8)}} = \frac{1}{{\sqrt { - \det \left( {{\cal G} + {\cal B}} \right)} }}\,\left( {{\gamma _{(8)}}\,\sigma _3^4 + {\gamma _{(6)}} \wedge {{\cal B}_{(2)}}\,\sigma _3^3 + \frac{1}{{2!}}{\gamma _{(4)}} \wedge {{\cal B}_{(2)}} \wedge {{\cal B}_{(2)}}\,\sigma _3^2} \right) \,i\,{\sigma _2}.
\end{equation}
After taking all the wedge products one finds
\begin{equation}
{({\Gamma _\kappa })_{(8)}} = \frac{{\left( {{M_8}\,\sigma _3^4 + {M_6}\,\sigma _3^3 + {M_4}\,\sigma _3^2} \right)\,i\,{\sigma _2}}}{{\sqrt { - \det \left( {{\cal G} + {\cal B}} \right)} }}\,d\tau  \wedge d\rho  \wedge d{\phi _1} \wedge d{\phi _2} \wedge d{\phi _3} \wedge d\alpha  \wedge d\beta  \wedge d\gamma \,.
\end{equation}
The Pauli matrices act on the two component Weyl spinor $\varepsilon  = \left( \begin{array}{l}
\varepsilon _1^{(\alpha)} \\
\varepsilon _2^{(\beta)}
\end{array} \right)$, $\alpha,\,\beta=1,\dots, 16$. The form of the matrices $M_4$, $M_6$ and $M_8$ is given in Appendix \ref{sec: Appendix-D}. After applying the Pauli matrices the kappa symmetry preserving condition (\ref{kappa symmetry preserving condition}) takes the form
\begin{equation}\label{D7 Kappa Symmetry System}
\frac{1}{{\sqrt { - \det \left( {{\cal G} + {\cal B}} \right)} }}\,\left( {{M_8}\,\left( \begin{array}{l}
\,\,\,\,\varepsilon _2^\beta \\
 - \varepsilon _1^\alpha
\end{array} \right) + {M_6}\,\left( \begin{array}{l}
\varepsilon _2^\beta \\
\varepsilon _1^\alpha
\end{array} \right) + {M_4}\,\left( \begin{array}{l}
\,\,\,\,\varepsilon _2^\beta \\
 - \varepsilon _1^\alpha
\end{array} \right)} \right) = \left( \begin{array}{l}
\varepsilon _1^\alpha \\
\varepsilon _2^\beta
\end{array} \right)\,.
\end{equation}
This is an algebraic system for the components of the spinor. We can solve it if we choose a simpler D7-brane embedding ansatz, namely
\begin{equation}
\theta  = \theta (\rho ),\quad \phi  = \phi (\beta )\,.
\end{equation}
In this case the system (\ref{D7 Kappa Symmetry System}) takes the form
\begin{equation}\label{D7 System1}
\sum\limits_{\beta  = 1}^{16} {{a_{\alpha \beta }}\,\epsilon _2^{(\beta )} = } \epsilon _1^{(\alpha )},\quad \alpha  = 1, \ldots ,16\,,
\end{equation}
\begin{equation}\label{D7 System2}
\sum\limits_{\beta  = 1}^{16} {{b_{\alpha \beta }}\,\epsilon _1^{(\beta )} = } \epsilon _2^{(\alpha )},\quad \alpha  = 1, \ldots ,16\,,
\end{equation}
where the coefficients $a_{\alpha\beta}$ and $b_{\alpha\beta}$ are functions of the coordinates. We can substitute the $\epsilon _2^{(\alpha )}$ from (\ref{D7 System2}) into (\ref{D7 System1}), so that we end up with a homogenous system of 16 equations for the components $\epsilon _1^{(\alpha )}$:
\begin{equation}
\sum\limits_{\beta  = 1}^{16} {{s_{\alpha \beta }}\,\varepsilon _1^{(\beta )} = } 0,\quad \alpha  = 1, \ldots ,16\,.
\end{equation}
A homogenous system has a non-trivial solution if the determinant of its matrix is zero. In our case we have to impose $\det (s_{\alpha \beta })=0$. This condition implies
\begin{equation}
\theta=0, \quad \phi=-\beta+c\,,
\end{equation}
and the solutions are only the constant spinors as shown in Appendix \ref{sec: Appendix-D}.
\subsection{Kappa symmetry matrix for the D5-brane}
\label{subsec3.2: D5 kappa symmetry matrix}
In order to find the D5 kappa symmetry matrix we consider the most general D5 embedding ansatz of the form
\begin{align}
{\xi ^a} = (\tau ,\rho ,{\phi _2},{\phi _3},\beta ,\gamma ),\quad {\phi _1} = {\phi _1}({\xi ^a}),\quad \theta  = \theta ({\xi ^a}),\quad \alpha  = \alpha ({\xi ^a}),\quad \phi  = \phi ({\xi ^a})\,.
\end{align}
The 6-form kappa symmetry matrix for the D5-brane, $p=5$, $k=2$, looks like
\begin{equation}
{({\Gamma _\kappa })_{(6)}} = \frac{1}{{\sqrt { - \det \left( {{\cal G} + {\cal B}} \right)} }}\,\left( {{\gamma _{(4)}} \wedge {{\cal B}_{(2)}}\,\sigma _3^2 + \frac{1}{{2!}}\,{\gamma _{(2)}} \wedge {{\cal B}_{(2)}} \wedge {{\cal B}_{(2)}}\,\sigma _3^1} \right){\mkern 1mu} \,i\,{\sigma _2}.
\end{equation}
After taking the wedge products one finds
\begin{equation}
{({\Gamma _\kappa })_{(6)}} = \frac{{{M_2}\,i\,\sigma _3^1\,{\sigma _2} + {M_4}\,i\,\sigma _3^2\,{\sigma _2}}}{{\sqrt { - \det \left( {{\cal G} + {\cal B}} \right)} }}\,d\tau  \wedge d\rho  \wedge d{\phi _2} \wedge d{\phi _3} \wedge d\beta  \wedge d\gamma \,.
\end{equation}
The explicit form of the matrices $M_2$ and $M_4$ is given in Appendix \ref{secC:Appendix-E}. Applying the Pauli matrices, the projection equation (\ref{kappa symmetry preserving condition}) becomes
\begin{equation}\label{D5 Kappa Symmetry System}
{M_2}\,\left( {\begin{array}{*{20}{l}}
{\varepsilon _2^\beta }\\
{\varepsilon _1^\alpha }
\end{array}} \right) + {M_4}\,\left( {\begin{array}{*{20}{l}}
{\,\,\,\varepsilon _2^\beta }\\
{ - \varepsilon _1^\alpha }
\end{array}} \right) = \sqrt { - \det \left( {{\cal G} + {\cal B}} \right)} \,\left( {\begin{array}{*{20}{l}}
{\varepsilon _1^\alpha }\\
{\varepsilon _2^\beta }
\end{array}} \right)\,.
\end{equation}
Considering the following simpler D5 embedding ansatz
\begin{equation}
{\phi _1} = \frac{\pi }{2}\,,\quad \theta  = \theta (\rho )\,,\quad \alpha  = \alpha ({\phi _2})\,,\quad \phi  = \phi (\beta )\,,
\end{equation}
one can show that the system (\ref{D5 Kappa Symmetry System}) has constant solutions if
\begin{equation}\label{D5KappaEmbedding}
  {\phi _1} = \frac{\pi }{2}\,,\quad \theta  = 0\,,\quad \alpha  = \frac{\pi }{2}\,,\quad \phi  =  - \beta  + c\,.
\end{equation}
This is also compatible with the D5-brane equations of motion.
\section{D7-brane scalar fluctuations and the meson spectrum}
\label{sec:D7-Scalar-Fluctuations}
In this section we consider the kappa symmetric $D7$-brane embedding and study the corresponding energy spectrum of the scalar fluctuations. We show that the ground state of the spectrum along the $\theta$ direction is not equal to the conformal dimension of the operators dual to the fluctuations.
\subsection{Fluctuations along $\phi$}
\label{subsec3-2:D7-Fluctuations-Along-Phi}
The D7 embedding ansatz
\begin{equation}\label{solD7KappasymetricAnsatz}
  \theta (\rho ) = const,\quad \phi (\beta ) =  - \beta {\rm{ + c}}
\end{equation}
satisfies both the kappa symmetry preserving condition and the classical D7-brane embedding equations. Next we are going to redefine the coordinate $\phi\to\phi+\beta$ in order to consider fluctuations around $\phi=const$. The fluctuation ansatz is given by
\begin{equation}
  \theta  = 0 + \eta \,\Theta ,\quad \phi  = 0 + \eta \,\Phi \,.
\end{equation}
Taking into account the proper shift in the metric we can expand the DBI and the WZ Lagrangians up to quadratic order in the fluctuations and keeping only terms of order $\eta^2=(2\,\pi\,\alpha')^2$ we find\footnote{For the DBI Lagrangian expansion we used the formula
\begin{equation*}
\sqrt {\det \left( {\hat 1 + \hat M} \right)}  = 1 + \frac{1}{2}\,{\mathop{\rm Tr}\nolimits} \hat M - \frac{1}{4}\,{\mathop{\rm Tr}\nolimits} {{\hat M}^2} + \frac{1}{8}\,{{\mathop{\rm Tr}\nolimits} ^2}\hat M - \frac{1}{8}\,{\mathop{\rm Tr}\nolimits} {{\hat M}^2}\,{\mathop{\rm Tr}\nolimits} \hat M + \frac{1}{{32}}{{\mathop{\rm Tr}\nolimits} ^2}{{\hat M}^2} +  \cdot  \cdot  \cdot \,.
\end{equation*}
It is derived from the following general formula
\begin{equation*}
\sqrt {\det \left( {\hat 1 + \hat M} \right)}  = {e^{\frac{1}{2}\,\ln \,\det \left( {\hat 1 + \hat M} \right)}} = {e^{\frac{1}{2}\,{\rm{Tr}}\ln \left( {\hat 1 + \hat M} \right)}} = {e^{\sum\limits_{j = 1}^\infty  {\frac{{{{( - 1)}^{j + 1}}{\kern 1pt} {\rm{Tr}}({{\hat M}^j})}}{{2j}}} }} = \sum\limits_{k = 0}^\infty  {\,\frac{1}{{k!}}\,{{\left( {\sum\limits_{j = 1}^\infty  {\frac{{{{( - 1)}^{j + 1}}\,{\rm{Tr}}\left( {{{\hat M}^j}} \right)}}{{2\,j}}} } \right)}^k}} \,.
\end{equation*}
}
\begin{equation}\label{sec3-2:D7-Quadratic-Lagrangian}
  {\cal L}_{DBI}^{(2)} =  - {\mu _7}\,\frac{{{\eta ^2}}}{2}\,\sqrt { - \det g} \,\left( {{G_{\phi \phi }}\,{g^{ac}}\,{\partial _a}\Phi \,{\partial _c}\Phi  + } \right.\left. {{G_{\theta \theta }}\,{g^{ac}}\,{\partial _a}\Theta \,{\partial _c}\Theta } \right)\,,
\end{equation}
where $g=\det(g_{ab})$ is the determinant of the induced metric, $G_{\phi \phi}$ and $G_{\theta \theta}$ are the metric components in front of $d\phi^2$ and $d\theta^2$. The WZ part takes the form
\begin{align}
{\cal L}_{WZ}^{(2)} = \,{h_0}\,\left( {{h_1}\,\Theta \,{\partial _\beta }\Phi  + {h_2}\,\Theta \,{\partial _\rho }\Theta  + {h_3}\,{\partial _\beta }\Phi \,{\partial _\rho }\Theta  + {h_4}\,{\Theta ^2}} \right)\,,
\end{align}
where
\[{h_0} = \frac{{{2^{1/3}}\,{L^8}\,{\eta ^2}\,{\mu _7}\,\sin \alpha \,{{\sin }^2}{\phi _1}\,\sin {\phi _2}\,{{\sinh }^4}\rho }}{{243\,{{\left( {\cos 2\theta  - 3} \right)}^3}}}\,,\]
\[{h_1} = 4\,{\cos ^2}\theta \,\left( {\cos 2\theta  - 3} \right)\,\left( {237 - 377\,\cos 2\theta  + 55\,\cos 4\theta  - 3\,\cos 6\theta } \right)\,\theta '(\rho )\,,\]
\begin{align*}
{h_2} &=  - 4\,{\cos ^2}\theta \,\left( {\cos 2\theta  - 3} \right)\,\left( { - 134 - 237\,\phi '(\beta ) + \cos 6\theta \,\left( {2 + 3\,\phi '(\beta )} \right)} \right.\\
 &- \left. {\cos 4\theta \,\left( {34 + 55\,\phi '(\beta )} \right) + \cos 2\theta \,\left( {214 + 377\,\phi '(\beta )} \right)} \right)\,,
\end{align*}
\[{h_3} = 4\,{\cos ^3}\theta \,{\left( {\cos 2\theta  - 3} \right)^2}\,\left( {53\,\sin \theta  - 3\,\sin 3\theta } \right)\,,\]
\begin{align*}
{h_4} &= \theta '(\rho )\,\sin 2\theta \,\left( {304 + 512\,\phi '(\beta ) + 2\,\cos 8\theta \,\left( {2 + 3\,\phi '(\beta )} \right)} \right.\\
 &- \left. {\cos 6\theta \,\left( {78 + 121\,\phi '(\beta )} \right) + \cos 4\theta \,\left( {556 + 906\,\phi '(\beta )} \right) - 3\,\cos 2\theta \,\left( {710 + 1277\,\phi '(\beta )} \right)} \right)\,.
\end{align*}
The fluctuations for the scalar field $\Phi(\xi^a)$ are governed by the equation
\begin{equation}\label{eq:sec3-2:Laplace-Beltrami-Equation-For-The-Scalar-Field-Phi}
{\nabla ^a}{\nabla _a}\Phi  + \left( {{\partial _a}{G_{\phi \phi }}} \right)\,{g^{ab}}\,{\partial _b}\Phi  = 0\,,
\end{equation}
which is a Klein-Gordon-like equation, where
\begin{equation}
  {\nabla ^a}{\nabla _a}\Phi  = \frac{1}{{\sqrt { - g} }}\,{\partial _a}\left( {\sqrt { - g} \,{g^{ab}}\,{\partial _b}\,\Phi } \right)\,.
\end{equation}
After expanding the covariant derivatives in eq. (\ref{eq:sec3-2:Laplace-Beltrami-Equation-For-The-Scalar-Field-Phi}) one finds
\begin{equation}\
- \partial _\tau ^2\Phi  + {\cosh ^2}\rho \,{{\tilde \Delta }_\rho }\Phi  + {\coth ^2}\rho \,{\Delta _{{\phi _i}}}\Phi \, + 3\,{\cosh ^2}\rho \,{{\tilde \Delta }_{{\alpha _i}}}\Phi  = 0\,,
\end{equation}
where $\phi_i=(\phi_1,\,\phi_2,\,\phi_3)$, $\alpha_i=(\alpha,\,\beta,\,\gamma)$, and
\begin{equation}\label{eq:sec3-2:Radial-Laplacian}
{{\tilde\Delta }_\rho }\Phi  = \partial _\rho ^2\Phi  + (3\,\coth \rho  + \tanh \rho )\,{\partial _\rho }\Phi \,,
\end{equation}
\begin{equation}\label{eq:sec3-2:Hyperspherical-Laplacian-Phi1-Phi2-Phi3}
{\Delta _{{\phi _i}}}\Phi  = \frac{{{\partial _{{\phi _1}}}\left( {{{\sin }^2}{\phi _1}\,{\partial _{{\phi _1}}}\Phi } \right)}}{{{{\sin }^2}{\phi _1}}} + \frac{{{\partial _{{\phi _2}}}\left( {\sin {\phi _2}\,{\partial _{{\phi _2}}}\Phi } \right)}}{{{{\sin }^2}{\phi _1}\,\sin {\phi _2}}} + \frac{{\partial _{{\phi _3}}^2\Phi }}{{{{\sin }^2}{\phi _1}\,{{\sin }^2}{\phi _2}}}\,,
\end{equation}
\begin{equation}\label{eq:sec3-2:Squashed-Laplacian-Alpha-Beta-Gamma}
{{ \tilde\Delta }_{{\alpha _i}}}\Phi  = \frac{1}{{\sin \alpha }}\,{\partial _\alpha }\left( {\sin \alpha \,{\partial _\alpha }\Phi } \right) + \frac{1}{{{{\sin }^2}\alpha }}\,\left( {\frac{{\cos 2\alpha + 7}}{8}\,\partial _\beta ^2\Phi  + \partial _\gamma ^2\Phi  - 2\,\cos \alpha \,{\partial^2_{\beta\gamma}}\Phi } \right)\,.
\end{equation}
Separation of variables of the form $\Phi=e^{i\,\omega\,\tau}\,R(\rho)\,\mathcal{Y}^{\ell}(S^{3}_{\phi_i})\,Z(\tilde{S}^{3}_{\alpha_i})$ leads to the following set of spectral equations:
\begin{equation}
\label{eq:sec3-2:Tau-Equation}
\ddot T(\tau ) =  - {\omega ^2}\,T(\tau )\,,
\end{equation}
\begin{equation}
\label{eq:sec3-2:Spectral-Equation-For-Nu}
{{\tilde \Delta }_{{\alpha _i}}}Z({\alpha _i}) =  - \nu \,Z({\alpha _i})\,,
\end{equation}
\begin{equation}
\label{eq:sec3-2:Hyperspherical-Laplacian-Phi1-Phi2-Phi3-Equation}
{\Delta _{{\phi _i}}}{\mathcal{Y}^{\ell}}({\phi _i}) =  -\ell\,(\ell + 2)\,{\mathcal{Y}^{\ell}}({\phi _i})\,,
\end{equation}
\begin{equation}
\label{eq:sec3-2:Radial-Equation-With-Variable-Rho}
R''(\rho ) + (3\,\coth \rho  + \tanh \rho )\,R'(\rho ) + \left( {\frac{{{\omega ^2}}}{{{{\cosh }^2}\rho }} - \frac{{\ell\,(\ell + 2)}}{{{{\sinh }^2}\rho }} - 3\,\nu } \right)\,R(\rho ) = 0\,,
\end{equation}
where $\omega$ is the energy of the fluctuations, $\nu$ is the eigenvalue of the operator (\ref{eq:sec3-2:Spectral-Equation-For-Nu}), and $\mathcal{Y}^{\ell}({\phi _i})$ are the hyperspherical harmonics, $\ell\in\mathbb{N}_0$. In order to facilitate the calculation of the conformal dimension $\Delta$ and the spectrum, we make the following change of variables $\sinh\rho=r$ in the radial equation (\ref{eq:sec3-2:Radial-Equation-With-Variable-Rho}):
\begin{equation}\label{eq:sec3-2:Radial-Equation-With-Variable-r}
 R''(r) + \frac{{3 + 5\,{r^2}}}{{r\,\left( {{r^2} + 1} \right)}}\,R'(r) + \left( {\frac{{{\omega ^2}}}{{{{\left( {{r^2} + 1} \right)}^2}}} - \frac{{\ell\,(\ell + 2)}}{{{r^2}\,\left( {{r^2} + 1} \right)}} - \frac{{3\,\nu }}{{{r^2} + 1}}} \right)\,R(r) = 0.
\end{equation}
There are two independent solutions given in terms of the standard Gaussian hypergeometric function:
\begin{equation}
R(r) = {R_ + }(r) + {R_ - }(r)\,,
\end{equation}
\begin{equation}
{R_ + }(r)={c_1}\,{r^{ - 2 - \ell}}{\left( {{r^2} + 1} \right)^{ - \frac{\omega }{2}}}{\,_2}{F_1}\left( {a,\,b;\,c;\,z} \right)\,,
\end{equation}
\begin{equation}
{R_ - }(r) =  {c_2}\,{r^\ell}\,{\left( {{r^2} + 1} \right)^{ - \frac{\omega }{2}}}{\,_2}{F_1}\left( {a - c + 1,\,b - c + 1;\,2 - c;\,z} \right)\,,
\end{equation}
with
\begin{equation*}
  a = \frac{1}{2}\,\left( { - \ell - \omega  - \sqrt {3\,\nu  + 4} } \right),\quad
  b = \frac{1}{2}\,\left( { - \ell - \omega  + \sqrt {3\,\nu  + 4} } \right),
  \quad c =  - \ell,
  \quad z =  - {r^2}\,.
\end{equation*}
The only regular solution at the origin $r=0$ is $R_{-}(r)$, which makes it our choice for normalizable solution. To assure normalizability at infinity one has to terminate the series of the hypergeometric function at some finite non-negative integer power $n$. As it is well known the hypergeometric function becomes a polynomial of degree $n$ if one of its first two arguments is a negative integer $-n$, $n\geq 0$. Therefore setting ${b - c + 1}=-n$ gives the quantization condition and the form of the scalar meson spectrum\footnote{There exists a symmetry between interchanging the first two arguments of the hypergeometric function, i.e.  $_{2}F_1(a,\,b;\,c;\,z)={_{2}}F_1(b,\,a;\,c;\,z)$, which means we have the freedom to choose which argument to quantize. In our case we chose to have positive energy without any loss of generality.}:
\begin{equation}
\omega  = \sqrt {4 + 3\,\nu }  + 2 + \ell + 2\,n\,.
\end{equation}
By the standard AdS/CFT dictionary one can calculate the conformal dimension of the operators corresponding to $\Phi$ from the analysis of the radial equation (\ref{eq:sec3-2:Radial-Equation-With-Variable-r}) at the boundary $r\to\infty$. The large $r$ behaviour is determined by the following asymptotic equation
\begin{equation}
 R''(r) + \frac{5}{r}\,R'(r) - \frac{{3\,\nu }}{{{r^2}}}\,R(r) = 0\,,
\end{equation}
with a general solution:
\begin{equation}
 R(r) = {c_1}\,{r^{ - \sqrt {3\,\nu  + 4}  - 2}} + {c_2}\,{r^{\sqrt {3\,\nu  + 4}  - 2}}\,.
\end{equation}
This solution contains normalizable and non-normalizable parts that behaves as $r^{k_{1}}=r^{\Delta-4+p}$ and $\ r^{k_{2}}=r^{-\Delta+p}$, for some constant $p$. According to the AdS/CFT dictionary taking the difference of the powers one finds the conformal dimension
\begin{equation}\label{eq:sec3-2:Conformal-Dimension-For-Phi-Fluctuation-Theta-Equal-Zero-Case}
  \Delta  = \frac{{{k_1} - {k_2}}}{2} + 2 = 2 + \sqrt {3\,\nu  + 4}\,,
\end{equation}
where ${k_1} = -2+\sqrt {3\,\nu  + 4}$, and ${k_2} =  -2- \sqrt {3\,\nu  + 4}.$ Equation (\ref{eq:sec3-2:Conformal-Dimension-For-Phi-Fluctuation-Theta-Equal-Zero-Case}) allows us to express the spectrum in terms of the conformal dimension:
\begin{equation}\label{eq:sec3-2:The-Meson-Spectrum}
  \omega =  \Delta  + \ell + 2\,n\,.
\end{equation}
We can conclude that the energy of the ground state is given by the conformal dimension of the operator dual to the fluctuations. This is what we expect for supersymmetric embeddings of the D7-brane, which is consistent with similar results for the meson spectra found in different background geometries \cite{FilevGlobal:2010, Filev:2013}. For higher modes the spectrum is equidistant.
\subsection{Fluctuations along $\theta$}
\label{subsec3-2:D7-Fluctuations-Along-Phi-For-Theta-Equal-Zero}
%
Now we proceed with the study of the D7-brane scalar fluctuations along the $\theta$ direction. The equation governing the fluctuations is given by
\begin{equation}\label{D7ThetaFluctuationEquation}
  - \partial _\tau ^2\Theta  + {\cosh ^2}\rho \,{{\tilde \Delta }_\rho }\Theta  + {\coth ^2}\rho \,{\Delta _{{\phi _i}}}\Theta  + 3\,{\cosh ^2}\rho \,{{\tilde \Delta }_{{\alpha _i}}}\Theta  + 3\,{\cosh ^2}\rho \,\Theta  = 0\,,
\end{equation}
where
\begin{equation}
  {{\tilde \Delta }_\rho }\Theta  = \partial _\rho ^2\Theta  + \left( {3\,\coth \rho  + \frac{5}{2}\,\tanh \rho } \right)\,{\partial _\rho }\Theta \,,
\end{equation}
\begin{equation}
  {\Delta _{{\phi _i}}}\Phi  = \frac{{{\partial _{{\phi _1}}}\left( {{{\sin }^2}{\phi _1}\,{\partial _{{\phi _1}}}\Phi } \right)}}{{{{\sin }^2}{\phi _1}}} + \frac{{{\partial _{{\phi _2}}}\left( {\sin {\phi _2}\,{\partial _{{\phi _2}}}\Phi } \right)}}{{{{\sin }^2}{\phi _1}\,\sin {\phi _2}}} + \frac{{\partial _{{\phi _3}}^2\Phi }}{{{{\sin }^2}{\phi _1}\,{{\sin }^2}{\phi _2}}}\,,
\end{equation}
\begin{equation}\label{D7OperatorForNuTheta}
{{\tilde \Delta }_{{\alpha _i}}}\Theta  = \frac{1}{{\sin \alpha }}\,{\partial _\alpha }\left( {\sin \alpha \,{\partial _\alpha }\Theta } \right) + \frac{1}{{{{\sin }^2}\alpha }}\,\left( {\frac{{\cos 2\alpha  + 7}}{{32}}\,\partial _\beta ^2\Theta  + \partial _\gamma ^2\Theta  - {\partial _{\beta \gamma }}\Theta \,\cos \alpha } \right){\mkern 1mu} \,.
\end{equation}
The form of eq. (\ref{D7ThetaFluctuationEquation}) assumes separation of variables $\Theta(\xi^a)=e^{i\,\omega\,\tau}\,R(\rho)\,\mathcal{Y}^{\ell}(S^{3}_{\phi_i})\,Z(\tilde{S}^{3}_{\alpha_i})$. This leads to the following radial equation:
\begin{equation}
R''(\rho ) + \left( {\frac{5}{2}\,\tanh \rho  + 3\,\coth \rho } \right)\,R'(\rho ) + \left( {\frac{{{\omega ^2}}}{{{{\cosh }^2}\rho }} - \frac{{\ell \,\left( {\ell  + 2} \right)}}{{{{\sinh }^2}\rho }} - 3\,\nu  + 3} \right)\,R(\rho ) = 0\,.
\end{equation}
where $\nu$ is the eigenvalue quantum number for the operator defined in (\ref{D7OperatorForNuTheta}). Changing the radial variable to $r=\sinh\rho$ we arrive at
\begin{equation}
R''(r ) + \frac{{6 + 13\,{r^2}}}{{2\,\left( {r + {r^3}} \right)}}\,R'(r ) + \left( {\frac{{{\omega ^2}}}{{{{\left( {1 + {r^2}} \right)}^2}}} - \frac{{\ell \,\left( {\ell  + 2} \right)}}{{{r^2}\,\left( {1 + {r^2}} \right)}} - \frac{{3\,\nu  - 3}}{{1 + {r^2}}}} \right)\,R(r ) = 0\,.
\end{equation}
Its regular solution is given by
\begin{equation}
  R(r) = {\left( {{r^2} + 1} \right)^{ - \frac{1}{8}\,\left( {3 + \sqrt {16\,{\omega ^2} + 9} } \right)}}\,{r^l}\,F\left( {a,\,b,\,\ell  + 2,\, - {r^2}} \right)\,,
\end{equation}
where
\begin{equation}
  a = \frac{1}{8}\,\left( {\sqrt {73 + 48\,\nu }  - \sqrt {16\,{\omega ^2} + 9} } \right) + \frac{\ell }{2} + 1,\quad b =  - \frac{1}{8}\,\left( {\sqrt {16\,{\omega ^2} + 9}  + \sqrt {73 + 48\,\nu } } \right) + \frac{\ell }{2} + 1\,.
\end{equation}
Normalizability of the solution implies restrictions from which one finds the the meson spectrum along $\theta$:
\begin{equation}
  {\omega ^2} = {\left( {\Delta  + \ell  + 2\,n} \right)^2} - \frac{9}{{16}}\,.
\end{equation}
The ground state is determined by
\begin{equation}
\omega _0^2 = {\Delta ^2} - \frac{9}{{16}}\,.
\end{equation}
where $\Delta = \left( {8 + \sqrt {73 + 48\,\nu } } \right)/4$ is the conformal dimension. One notes that an additional shift is present in the ground state. Its existence may be due to the fact that we are considering only embeddings compatible with the kappa symmetry, rather than supersymmetry. This result differs from the results found in recent papers \cite{FilevGlobal:2010, Filev:2013}. A complete study of the kappa symmetry condition may uncover D7-brane embeddings with a vanishing shift. If not it may give us clues for its origin.
\section{D5 probe brane scalar fluctuations and mesons}
\label{sec:D5-Scalar-Fluctuations}
In this section we consider the kappa symmetric D5 probe brane embedding and analyse the corresponding spectrum of the scalar fluctuations. The considerations conceptually repeats those in the previous section.
\subsection{Fluctuations along $\alpha,\,\theta$ and $\phi$}
\label{subsec4-2:D5-Fluctuations-Along-Phi-For-Theta-Equal-Zero}
In order to find the fluctuations of the D5-brane we use the embedding from subsection \ref{subsec3.2: D5 kappa symmetry matrix} and the pullbacks of the fields as (\ref{Pullback of the PW metric in static gauge}) and (\ref{Pullback of the RR potential}). One can make the convenient shift $\phi\to\phi+\beta=const$ in eq. (\ref{D5KappaEmbedding}) to consider only constant fluctuations along $\phi$. Therefore the D5 probe brane fluctuation ansatz can be written by
\begin{equation}
{\phi _1} = \frac{\pi }{2} + \eta \,{\Phi _1},\quad \theta  = 0 + \eta \,\Theta ,\quad \alpha  = \frac{\pi }{2} + \eta \,{\cal A},\quad \phi  = 0 + \eta \,\Phi\,,
\end{equation}
where $\phi_1,\,\alpha,\,\theta,\,\phi$ are normal directions to the brane's world volume, and $\eta=2\,\pi\,\alpha'$. In this case the fluctuation equations for $\mathcal{A},\,\Theta$ and $\Phi$ have the same form
\begin{equation}\label{eq:sec4-2:D5-Fluctuations-Along-Phi-For-Theta-Equal-Zero}
  - \partial _\tau ^2{\cal P} + {\cosh ^2}\rho \,{{\tilde \Delta }_\rho }{\cal P} + {\coth ^{\rm{2}}}\rho \,{\Delta _{{\phi _2}{\phi _3}}}{\cal P} + 3\,{\cosh ^{\rm{2}}}\rho \,{{\tilde \Delta }_{\beta \gamma }}{\cal P} = 0\,,
\end{equation}
where $\mathcal{P}$ stands for any of the scalar fields $\mathcal{A},\,\Theta$ or $\Phi$. The differential operators in eq. (\ref{eq:sec4-2:D5-Fluctuations-Along-Phi-For-Theta-Equal-Zero}) are given by
\begin{equation}\label{D5DeltaBetaGamma}
{{\tilde \Delta }_{\beta \gamma }}{\cal P}  = \frac{3}{4}\,\partial _\beta ^2{\cal P}  + \partial _\gamma ^2{\cal P}\,,
\end{equation}
\begin{equation}
{{\tilde \Delta }_\rho }{\cal P}  = \partial _\rho ^2{\cal P}  + \left( {2\,\coth \rho  + \tanh \rho } \right)\,{\partial _\rho }{\cal P} \,,
\end{equation}
\begin{equation}\label{D5DeltaPhis}
{\Delta _{{\phi _2}{\phi _3}}}{\cal P}  = \frac{1}{{\sin {\phi _2}}}\,{\partial _{{\phi _2}}}(\sin {\phi _2}\,{\partial _{{\phi _2}}}{\cal P} ) + \frac{1}{{{{\sin }^2}{\phi _2}}}\,\partial _{{\phi _3}}^2{\cal P} \,.
\end{equation}
Separation of variables, ${\cal P}=e^{i\,\omega\,\tau}\,R(\rho)\,\mathcal{Y}^{\ell}(\phi_2,\,\phi_3)\,Z(\beta,\,\gamma)$, leads to the following spectral equations:
\begin{equation}
  \ddot T(\tau ) =  - {\omega ^2}\,T(\tau ),\quad
  \frac{{{\Delta _{{\phi _2}{\phi _3}}}{\mathcal{Y}^{\ell}}({\phi _2},{\phi _3})}}{{{\mathcal{Y}^{\ell}}({\phi _2},{\phi _3})}} = -\ell\,(\ell + 1),\quad
  \frac{{{{\tilde \Delta }_{\beta \gamma }}Z(\beta ,\gamma )}}{{Z(\beta ,\gamma )}} = -\nu \,,
\end{equation}
\begin{equation}\label{eq:sec4-2:D5-Brane-Radial-Equation-With-Variable-Rho}
  R''(\rho ) + (2\,\coth \rho  + \tanh \rho )\,R'(\rho ) + \left( {\frac{{{\omega ^2}}}{{{{\cosh }^2}\rho }} - \frac{{\ell\,(\ell + 1)}}{{{{\sinh }^2}\rho }} - 3\,\nu } \right)\, R(\rho ) = 0\,.
\end{equation}
It is more convenient to change the radial coordinate to $r=\sinh \rho$, after which equation (\ref{eq:sec4-2:D5-Brane-Radial-Equation-With-Variable-Rho}) takes the form
\begin{equation}
R''(r) + \frac{{2 + 4\,{r^2}}}{{r\,\left( {{r^2} + 1} \right)}}\,R'(r) + \left( {\frac{{{\omega ^2}}}{{{{\left( {{r^2} + 1} \right)}^2}}} - \frac{{\ell\,(\ell + 1)}}{{{r^2}\,\left( {{r^2} + 1} \right)}} - \frac{{3\,\nu }}{{{r^2} + 1}}} \right)\, R(r) = 0\,.
\end{equation}
The regular solution to this equation is given by
\begin{equation}
  R(r) = {r^\ell}\,{\left( {{r^2} + 1} \right)^{ - \frac{\omega }{2}}}{\,_2}{F_1}\left( {a,\,b;\,\ell + \frac{3}{2};\, - {r^2}} \right)\,,
\end{equation}
where
\[a = \frac{1}{4}\,\left( {2\,\ell - \sqrt {12\,\nu  + 9}  - 2\,\omega  + 3} \right),\quad
b = \frac{1}{4}\,\left( {2\,\ell + \sqrt {12\,\nu  + 9}  - 2\,\omega  + 3} \right)\,.\]
Imposing normalizability $b=-n$ one finds for the meson spectrum:
\begin{equation}\label{eq:sec4-2:D5-Brane-Meson-Spectrum-For-Phi-Theta-Equal-Zero}
  \omega  = \frac{1}{2}\,\sqrt {12\,\nu  + 9}  + \ell + \frac{3}{2} + 2\,n\,.
\end{equation}
The solution near the boundary $r\to \infty$ is determined by the asymptotic equation
\begin{equation}
 R''(r) + \frac{4}{r}\,R'(r) - \frac{{3\,\nu }}{{{r^2}}}\,R(r) = 0\,.
\end{equation}
with the following solution
\begin{equation}\label{eq:sec4-2:D5-Brane-Asymptotic-Radial-Equation-With-Variable-r}
 R(r) = {c_1}\,{r^{\frac{1}{2}\,\left( { - \sqrt {12\,\nu  + 9}  - 3} \right)}} + {c_2}\,{r^{\frac{1}{2}\,\left( {\sqrt {12\,\nu  + 9}  - 3} \right)}} = {c_1}\,{r^{{k_2}}} + {c_2}\,{r^{{k_1}}}\,,
\end{equation}
Therefore one can calculate the conformal dimension
\begin{equation}
 \Delta  = 2 + \frac{{{k_1} - {k_2}}}{2} = 2 + \frac{1}{2}\,\sqrt {12\,\nu  + 9}\,.
\end{equation}
Finally, the spectrum (\ref{eq:sec4-2:D5-Brane-Meson-Spectrum-For-Phi-Theta-Equal-Zero}) takes the form
\begin{equation}\label{eq:sec4-2:The-Spectrum}
 \omega  = \Delta - \frac{1}{2} + \ell + 2\,n\,.
\end{equation}
The ground state ($n,\,\ell=0$) is not equal to the conformal dimension $\Delta$, but it is shifted by --1/2, which implies once again that the chosen kappa symmetric D5 embedding may not be compatible with supersymmetry. Another origin of this shift is also not ruled out.
\subsection{Fluctuations along $\phi_1$}
The equation describing the scalar fluctuations along the $\phi_1$ direction is given by
\begin{equation}
- \partial _\tau ^2{\Phi _1} + {\cosh ^2}\rho \,{{\hat \Delta }_\rho }{\Phi _1} + {\coth ^{\rm{2}}}\rho \,{\Delta _{{\phi _2}{\phi _3}}}{\Phi _1} + 3\,{\cosh ^{\rm{2}}}\rho \,{{\tilde \Delta }_{\beta \gamma }}{\Phi _1} = 0\,,
\end{equation}
where
\begin{equation}
{{\hat \Delta }_\rho }{\Phi _1} = \partial _\rho ^2{\Phi _1} + \left( {4\,\coth \rho  + \tanh \rho } \right)\,{\partial _\rho }{\Phi _1}\,.
\end{equation}
The operators ${{\tilde \Delta }_{\beta \gamma }}{\Phi _1}$ and ${\Delta _{{\phi _2}{\phi _3}}}{\Phi _1}$ are defined as in (\ref{D5DeltaBetaGamma}) and (\ref{D5DeltaPhis}) respectively. The radial equation here case takes the form
\begin{equation}\label{eq:sec4-2:D5-Brane-Radial-Equation-With-Variable-Rho1}
  R''(\rho ) + (4\,\coth \rho  + \tanh \rho )\,R'(\rho ) + \left( {\frac{{{\omega ^2}}}{{{{\cosh }^2}\rho }} - \frac{{\ell\,(\ell + 1)}}{{{{\sinh }^2}\rho }} - 3\,\nu } \right)\, R(\rho ) = 0\,.
\end{equation}
Changing the radial variable again to $\sinh \rho  = r$ one finds
\begin{equation}
R''(r) + \frac{{4 +6\,{r^2}}}{r\,{\left( {{r^2} + 1} \right)}}\,R'(r) + \left( {\frac{{{\omega ^2}}}{{{{\left( {{r^2} + 1} \right)}^2}}} - \frac{{\ell\,(\ell + 1)}}{{{r^2}\,\left( {{r^2} + 1} \right)}} - \frac{{3\,\nu }}{{{r^2} + 1}}} \right){\mkern 1mu} R(r) = 0\,.
\end{equation}
The regular solution is written by
\begin{equation}
R(r) = {r^{\frac{1}{2}\,\left( {\sqrt {4\,\ell \,(\ell  + 1) + 9}  - 3} \right)}}\,{\left( {{r^2} + 1} \right)^{ - \frac{\omega }{2}}}{\,_2}{F_1}\left( {a,\,b,\,c,\, - {r^2}} \right)\,,
\end{equation}
where
\[a = \frac{1}{4}\,\left( { - 2\,\omega  + \sqrt {4\,\ell\,(\ell + 1) + 9}  - \sqrt {12\,\nu  + 25}  + 2} \right),\]
\[b = \frac{1}{4}\,\left( { - 2\,\omega  + \sqrt {4\,\ell\,(\ell + 1) + 9}  + \sqrt {12\,\nu  + 25}  + 2} \right)\,,\]
\[
c = \frac{1}{2}\,\left( {\sqrt {4\,\ell \,(\ell  + 1) + 9}  + 2} \right)\,.
\]
Imposing normalizability, $b=-n$, one finds the form of the meson spectrum
\begin{equation}\label{eq:sec4-2:D5-Brane-Meson-Spectrum-For-Phi-Theta-Equal-Zero1}
 \omega  = \Delta  - 1 + \frac{1}{2}\,\sqrt {4\,\ell\,(\ell + 1) + 9}  + 2\,n\,,
\end{equation}
where $\Delta  = \frac{1}{2}\,\sqrt {12\,\nu  + 25}  + 2$, and a ground state given by
\begin{equation}
 \omega_0=\Delta+\frac{1}{2}\,.
\end{equation}
Once again we observe that the ground state of the spectrum is shifted.
\section{Conclusion}
\label{sec:Conclusion}
\indent Quantum chromodynamics is the most successful theory describing the strong nuclear force so far. In the low energy regime of the theory the usual perturbative techniques are not applicable, which forces us to look for alternative non-perturbative methods. Such alternative techniques arise in string theory in the context of the AdS/CFT correspondence, where the physics of the supersymmetric Yang-Mills systems -- the theory giving the best approach to QCD -- can be understood by that of the D-brane dynamics and vice versa. Considering D-brane embeddings of various dimensionality in Pilch-Warner background is important due to the fact that the holographic dual of PW background is the non-trivial $\N=1$ supersymmetric fixed point of $\N=4$ Yang-Mills theory. The significantly reduced supersymmetry of the dual theory can bring the duality closer to realistic QCD-like models.
\\
\indent In this study we considered the D5- and D7-brane embeddings compatible with the kappa symmetry preserving condition and the brane equations of motion. Working in the case when the number $N_f$ of the flavour branes is much smaller than the number of the color branes $N_c$ allows to neglect the backreaction of the background. The analysis of the scalar fluctuations of the probe branes and the derivation of the spectra are given analytically.
\\
\indent  We found the explicit form of the kappa symmetry matrix for both the D5 and D7 probe branes. By solving the kappa symmetry preserving condition we were able to find brane embeddings compatible with the kappa symmetry and the classical equations of motion. We were also able to prove that the kappa symmetry preserved exactly the half of the spinor degrees of freedom for both D5 and D7.
\\
\indent We considered the kappa symmetric D7-brane embedding and study the corresponding spectrum of scalar fluctuations. We showed that the ground state of
the spectrum along the $\phi$ direction is equal to the conformal dimension. However, an unexpected shift appeared in the ground state of the spectrum along the $\theta$ direction. The consideration of both kappa symmetric and supersymmetric brane embeddings may cancel the shift, but another reason for its existence should not be ruled out.
\\
\indent Finally, we analyzed the spectrum of the D5-brane fluctuations. Once again we considered a kappa symmetric D5 embedding. This time shifts appeared in the ground states of the spectra along all of the transverse directions. The origin of the shifts in ground states remains unclear.
\\
\indent All obtained spectra are equidistant, but some of their ground states are not equal to the conformal dimension of the operators dual to the fluctuations due to the presence of unexpected shifts. This result differs from the one already found in \cite{FilevGlobal:2010} and \cite{Filev:2013}. A reason for the existence of the shifts may be that the kappa symmetric brane embeddings may not be compatible with the supersymmetry. To rule out this one has to conduct a complete study of the symmetries. In any case the resolution of this puzzle is an interesting work.
\\
\indent It is known for a long time \cite{Buchel:2001} that the \PW supergravity is dual to the $\N=4$ SYM softly broken down to $\N=2$. In general adding masses to all chiral multiplets in $\N=4$ SYM breaks the SUSY to $\N=1$. However it can be enhanced to $\N=2$. The enhancement locus in the dual geometry lies at $\theta=\pi/2$. Recent studies \cite{Zarembo:2015, Zarembo:2014, Zarembo2:2014} present interesting results and uncover the nice structure on both sides of the AdS/CFT correspondence in this setup. It would be interesting to conduct the above analysis to this case. Initial considerations show that the equations for the fluctuations are entangled, so more thorough analysis is needed. Work on those issues is in progress and we hope to report on it soon.
\acknowledgments
The authors would like to thank V. Filev for collaboration at early stage of this
project. We thank H. Dimov for insightful discussions and V. Filev for careful reading of the draft. This work was partially supported by the bulgarian NSF grant DFNI T02/6.
\appendix
\section{Explicit form of the R-R and NS-NS potentials}
\label{secA:Appendix-A}
The dilaton/axion system is trivial along the flow, i.e. $C_0=C_8=0$. The $C_2$ potential takes the following form
\begin{equation}
\begin{split}
{{\rm{C}}_2} = \Re e({A_2}) &= {C_{\alpha \beta }}\,{\rm{d}}\alpha  \wedge {\rm{d}}\beta  + {C_{\alpha \gamma }}\,{\rm{d}}\alpha  \wedge {\rm{d}}\gamma  + {C_{\alpha \phi }}\,{\rm{d}}\alpha  \wedge {\rm{d}}\phi+ \\
 &+ {C_{\beta \gamma }}\,{\rm{d}}\beta  \wedge {\rm{d}}\gamma  + {C_{\gamma \phi }}\,{\rm{d}}\gamma  \wedge {\rm{d}}\phi  + {C_{\theta \alpha }}\,{\rm{d}}\theta  \wedge {\rm{d}}\alpha  + {C_{\theta \gamma }}\,{\rm{d}}\theta  \wedge {\rm{d}}\gamma\,,
\end{split}
\end{equation}
where its components in global PW coordinates are given by
\[\begin{array}{l}
{C_{\alpha \beta }} = {K}\,\sin \beta \,\sin 2\theta \,,\,
{C_{\alpha \gamma }} = {K}\,\cos \alpha \,\sin \beta \,\sin 2\theta \,,\,
{C_{\alpha \phi }} = 2\,{K}\,\sin \beta \,\sin 2\theta \,,\\
{C_{\beta \gamma }} = {K}\,\sin \alpha \,\cos \beta \,\sin 2\theta \,,\,
{C_{\gamma \phi }} =  - 2\,{K}\,\sin \alpha \,\cos \,\beta \,\sin \,2\theta \,,\,
{C_{\theta \alpha }} = {K}\,(\cos 2\theta  - 3)\,\cos \beta ,\\
{C_{\theta \gamma }} = {K}\,(\cos 2\theta - 3)\,\sin \alpha \,\sin \beta \,,\,
{K} =  - \frac{{2\,\sqrt[3]{2}\,{L^2}\,\cos \theta }}{{9\,(\cos 2\theta  - 3)}}\,.
\end{array}\]
The NS two-form
\begin{equation}\label{NSKalbRamondBField}
\begin{split}
{{\rm{B}}_2} = \Im m({A_2}) &= {B_{\alpha \beta }}\,{\rm{d}}\alpha  \wedge {\rm{d}}\beta  + {B_{\alpha \gamma }}\,{\rm{d}}\alpha  \wedge {\rm{d}}\gamma  + {B_{\alpha \phi }}\,{\rm{d}}\alpha  \wedge {\rm{d}}\phi + \\
 &+ {B_{\beta \gamma }}\,{\rm{d}}\beta  \wedge {\rm{d}}\gamma  + {B_{\gamma \phi }}\,{\rm{d}}\gamma  \wedge {\rm{d}}\phi  + {B_{\theta \alpha }}\,{\rm{d}}\theta  \wedge {\rm{d}}\alpha  + {B_{\theta \gamma }}\,{\rm{d}}\theta  \wedge {\rm{d}}\gamma
\end{split}
\end{equation}
has components
\[\begin{array}{l}
{B_{\alpha \beta }} = {B_0}\,\cos \beta \,\sin 2\theta \,,\,
{B_{\alpha \gamma }} = {B_0}\,\cos \alpha \,\cos \beta \,\sin 2\theta \,,\,
{B_{\alpha \phi }} = 2\,{B_0}\,\cos \beta \,\sin 2\theta \,,\\
{B_{\beta \gamma }} =  - {B_0}\,\sin \alpha \,\sin \beta \,\sin 2\theta ,\,
{B_{\gamma \phi }} = 2\,{B_0}\,\sin \alpha \,\sin \beta \,\sin 2\theta \,,\,
{B_{\theta \alpha }} =  - {B_0}\,(\cos 2\theta  - 3)\,\sin \beta \,,\\
{B_{\theta \gamma }} = {B_0}\,(\cos 2\theta  - 3)\sin \alpha \,\cos \beta \,,
{B_0} =  - {K}\,.
\end{array}\]
We also have expressions for the four-form potential $C_4$:
\begin{equation}
    {C_4} = \frac{{{2^{5/3}}\,{L^4}\,{{\sinh }^4}\rho \,{{\sin }^2}{\phi _1}\,\sin {\phi _2}}}{3}\,{\rm{d}}\tau  \wedge {\rm{d}}{\phi _1} \wedge {\rm{d}}{\phi _2} \wedge {\rm{d}}{\phi _3}\,,
\end{equation}
and its dual 4-form:
\begin{equation}
    {\tilde C_4} =  - \frac{{{2^{5/3}}\,{L^4}\,\left( {\cos 4\theta  - 6\,\cos 2\theta  + 16\,(\cos 2\theta  - 3)\,\ln (3 - \cos 2\theta ) - 31} \right)\,\sin \alpha }}{{81\,(\cos 2\theta  - 3)}}\, {\rm{d}}\alpha  \wedge {\rm{d}}\beta  \wedge {\rm{d}}\gamma  \wedge {\rm{d}}\phi \,.
\end{equation}
The six-form potential $C_6$ has more complicated structure given by
\begin{align}
{C_6} &=  - \frac{{16\,{L^6}\,{\varepsilon _\phi }\,{{\sinh }^4}\rho \,\sin 2\theta \,\cos \theta \,\cos \beta }}{{27\,(\cos 2\theta  - 3)}}\,d\tau  \wedge d{\phi _1} \wedge d{\phi _2} \wedge d{\phi _3} \wedge d\alpha  \wedge d\phi \nonumber\\
 &- \frac{{16\,{L^6}\,{\varepsilon _\phi }\,{{\sinh }^4}\rho \,\sin 2\theta \,\cos \theta \,\sin \alpha \,\sin \beta }}{{27\,(\cos 2\theta  - 3)}}\,d\tau  \wedge d{\phi _1} \wedge d{\phi _2} \wedge d{\phi _3} \wedge d\gamma  \wedge d\phi \nonumber\\
 &- \frac{{{L^6}\,{\varepsilon _\phi }}}{9}\,{\sinh ^4}\rho \,\cos \theta \,(\cos 2\theta  - 3)\,\sin \beta \,d\tau  \wedge d{\phi _1} \wedge d{\phi _2} \wedge d{\phi _3} \wedge d\theta  \wedge d\alpha \nonumber\\
 &+ \frac{{{L^6}\,{\varepsilon _\phi }}}{9}\,{\sinh ^4}\rho \,\cos \theta \,(\cos 2\theta  - 3)\,\sin \alpha \,\cos \beta \,d\tau  \wedge d{\phi _1} \wedge d{\phi _2} \wedge d{\phi _3} \wedge d\theta  \wedge d\gamma \nonumber\\
 &+ \frac{{{L^6}\,{\varepsilon _\phi }\,{{\sinh }^4}\rho \,\sin 2\theta \,\cos \theta \,(\cos 2\theta  - 11)\,\cos \beta }}{{27\,(\cos 2\theta  - 3)}}\,d\tau  \wedge d{\phi _1} \wedge d{\phi _2} \wedge d{\phi _3} \wedge d\alpha  \wedge d\beta \nonumber\\
 &+ \frac{{{L^6}\,{\varepsilon _\phi }\,{{\sinh }^4}\rho \,\sin 2\theta \,\cos \theta \,(\cos 2\theta  - 11)\,\cos \alpha \,\cos \beta }}{{27\,(\cos 2\theta  - 3)}}\,d\tau  \wedge d{\phi _1} \wedge d{\phi _2} \wedge d{\phi _3} \wedge d\alpha  \wedge d\gamma\nonumber\\
 &- \frac{{{L^6}\,{\varepsilon _\phi }\,{{\sinh }^4}\rho \,\sin 2\theta \,\cos \theta \,(\cos 2\theta  - 11)\,\sin \alpha \,\sin \beta }}{{27\,(\cos 2\theta  - 3)}}\,d\tau  \wedge d{\phi _1} \wedge d{\phi _2} \wedge d{\phi _3} \wedge d\beta  \wedge d\gamma\,,
\end{align}
where ${\varepsilon _\phi } = {\sin ^2}{\phi _1}\,\sin {\phi _2}$. All the R-R and NS-NS potentials are written explicitly in global coordinates and satisfy the corresponding Bianchi identities and equations of motion (\ref{Bianchi-Identity-For-The-R-R-Forms}).
\section{D=10 Dirac gamma matrices}
\label{secB:Appendix-B}
In order to calculate the kappa symmetry matrix for the corresponding D-branes one has to chose an appropriate basis for the 10-dimensional Dirac gamma matrices $\Gamma_\mu$. We choose to work with the Majorana-Weyl representation:
\[\begin{array}{l}
{\Gamma _0} = i\,{\sigma _1} \otimes 1 \otimes 1 \otimes 1 \otimes 1\,,\quad {\Gamma _1} = i\,\chi  \otimes \chi  \otimes \chi  \otimes \chi  \otimes \chi\,,\\
{\Gamma _2} = i\,\chi  \otimes \chi \otimes 1 \otimes {\sigma _1} \otimes \chi \,,\quad {\Gamma _3} = i\,\chi  \otimes \chi \otimes 1 \otimes {\sigma _3} \otimes \chi \,,\\
{\Gamma _4} = i\,\chi  \otimes \chi  \otimes {\sigma _1} \otimes \chi  \otimes 1\,,\quad {\Gamma _5} = i\,\chi  \otimes \chi \otimes {\sigma _3} \otimes \chi  \otimes 1\,,\\
{\Gamma _6} = i\,\chi  \otimes \chi \otimes \chi  \otimes 1 \otimes {\sigma _1}\,,\quad {\Gamma _7} = i\,\chi  \otimes \chi  \otimes \chi  \otimes 1 \otimes {\sigma _3}\,,\\
{\Gamma _8} = i\,\chi  \otimes {\sigma _1} \otimes 1 \otimes 1 \otimes 1\,,\quad {\Gamma _9} = i\,\chi  \otimes {\sigma _3} \otimes 1 \otimes 1 \otimes 1\,,
\end{array}\]
\[
{\Gamma _*} = {\sigma _3} \otimes 1 \otimes 1 \otimes 1 \otimes 1 = {\Gamma _0}\,{\Gamma _1}\,{\Gamma _2}\,{\Gamma _3}\,{\Gamma _4}\,{\Gamma _5}\,{\Gamma _6}\,{\Gamma _7}\,{\Gamma _8}\,{\Gamma _9} = \left( {\begin{array}{*{20}{c}}
{{1_{16 \times 16}}}&0\\
0&{ - {1_{16 \times 16}}}
\end{array}} \right)\,,
\]
where $\chi  = i\,{\sigma _2}$, and $\sigma_i$ are the Pauli matrices:
\begin{equation}
{\sigma ^0} = \left( {\begin{array}{*{20}{c}}
1&0\\
0&1
\end{array}} \right)\,,\quad {\sigma ^1} = \left( {\begin{array}{*{20}{c}}
0&1\\
1&0
\end{array}} \right)\,,\quad {\sigma ^2} = \left( {\begin{array}{*{20}{c}}
0&{ - i}\\
i&0
\end{array}} \right)\,,\quad {\sigma ^3} = \left( {\begin{array}{*{20}{c}}
1&0\\
0&{ - 1}
\end{array}} \right)\,.
\end{equation}
All the gamma matrices satisfy the Clifford algebra
\begin{equation}\label{a}
  \left\{ {{\Gamma _\mu},\,{\Gamma _\nu}} \right\} = 2\,{\eta _{\mu\nu}}\,{1_{32 \times 32}}\,,\quad \eta _{\mu\nu}=\diag{(-1,1,\dots,1)}\,.
\end{equation}
The rule for the tensor product is the one used by Mathematica (the right matrix goes into the left one):
\[\begin{array}{l}
a \otimes b = \left( {\begin{array}{*{20}{c}}
{{a_{11}}}&{{a_{12}}}\\
{{a_{21}}}&{{a_{22}}}
\end{array}} \right) \otimes \left( {\begin{array}{*{20}{c}}
{{b_{11}}}&{{b_{12}}}\\
{{b_{21}}}&{{b_{22}}}
\end{array}} \right) =
\left( {\begin{array}{*{20}{c}}
{\begin{array}{*{20}{c}}
{{a_{11}}\,{b_{11}}}&{{a_{11}}\,{b_{12}}}\\
{{a_{11}}\,{b_{21}}}&{{a_{11}}\,{b_{22}}}
\end{array}}&{\begin{array}{*{20}{c}}
{{a_{12}}\,{b_{11}}}&{{a_{12}}\,{b_{12}}}\\
{{a_{12}}\,{b_{21}}}&{{a_{12}}\,{b_{22}}}
\end{array}}\\
{\begin{array}{*{20}{c}}
{{a_{21}}\,{b_{11}}}&{{a_{21}}\,{b_{12}}}\\
{{a_{21}}\,{b_{21}}}&{{a_{21}}\,{b_{22}}}
\end{array}}&{\begin{array}{*{20}{c}}
{{a_{22}}\,{b_{11}}}&{{a_{22}}\,{b_{12}}}\\
{{a_{22}}\,{b_{21}}}&{{a_{22}}\,{b_{22}}}
\end{array}}
\end{array}} \right)\,.
\end{array}\]
This assures the correct form of $\Gamma_{*}=\diag{(1,\dots,1,-1,\dots,-1)}$.
\section{Vielbein coefficients}
\label{secC:Appendix-C}
The induced gamma matrices on the world volume are given by
\begin{equation}
  {\gamma _a} = {\partial _a}{X^M}\,E_M^I\,{\Gamma _I}\,,
\end{equation}
where $E_M^I$, $I=0,\dots,9$, are the vielbein coefficients of the PW metric
\begin{equation}\label{a}
  ds^2=\eta_{IJ}\,e^{I}\,e^{J}\,.
\end{equation}
The basis one-forms $ e^I=E^I_M\,dX^M$ take the form:
\[\begin{array}{l}
{e^0} = i\,L\,\Omega \,\cosh \rho \,{\rm{d}}\tau  = E_0^0\,{\rm{d}}\tau \,,\\
{e^1} = L\,\Omega \,{\rm{d}}\rho  = E_1^1\,{\rm{d}}\rho \,,\\
{e^2} = L\,\Omega \,\sinh \rho \,d\phi _1^{} = E_2^2\,d\phi _1^{}\,,\\
{e^3} = L\,\Omega \,\sinh \rho \,\sin {\phi _1}\,d\phi _2^{} = E_3^3\,d\phi _2^{}\,,\\
{e^4} = L\,\Omega \,\sinh \rho \,\sin {\phi _1}\,\sin {\phi _2}\,d\phi _3^{} = E_4^4\,d\phi _3^{}\,,\\
{e^5} = \sqrt {\frac{2}{3}} \,L\,\Omega \,{\rm{d}}\theta  = E_5^5\,{\rm{d}}\theta ,\\
{e^6} = \sqrt {\frac{2}{3}} \,L\,\Omega \,\frac{{2\,\cos \theta }}{{\sqrt {3 - \cos 2\theta } }}\,\sigma _1^{} = E_6^6\,{\rm{d}}\alpha  + E_8^6\,{\rm{d}}\gamma ,\\
{e^7} = \sqrt {\frac{2}{3}} \,L\,\Omega \,\frac{{2\,\cos \theta }}{{\sqrt {3 - \cos 2\theta } }}\,\sigma _2^{} = E_6^7\,{\rm{d}}\alpha  + E_8^7\,{\rm{d}}\gamma ,\\
{e^8} = \sqrt {\frac{2}{3}} L\,\Omega \,\frac{{2\,\sin 2\theta }}{{(3 - \cos 2\theta )}}\,({\sigma _3} + {\rm{d}}\phi ) = E_7^8\,{\rm{d}}\beta  + E_8^8\,{\rm{d}}\gamma  + E_9^8\,{\rm{d}}\phi \,,\\
{e^9} = \frac{2}{3}\,L\,\Omega \,\left( {\frac{{1 - 3\,\cos 2\theta }}{{\cos 2\theta  - 3}}} \right)\,\left( {{\rm{d}}\phi  - \frac{{4\,{{\cos }^2}\theta }}{{1 - 3\,\cos 2\theta }}\,{\sigma _3}} \right) = E_7^9\,{\rm{d}}\beta  + E_8^9\,{\rm{d}}\gamma  + E_9^9\,{\rm{d}}\phi \,,
\end{array}\]
where
\[\begin{array}{l}
{\sigma _1} = \frac{1}{2}\,(\sin \beta \,{\rm{d}}\alpha  - \cos \beta \,{\rm{sin}}\alpha \,{\rm{d}}\gamma )\,,\\
{\sigma _2} =  - \frac{1}{2}\,(\cos \beta \,{\rm{d}}\alpha  + \sin \beta \,{\rm{sin}}\alpha \,{\rm{d}}\gamma )\,,\\
{\sigma _3} = \frac{1}{2}\,({\rm{d}}\beta  + \cos \alpha \,{\rm{d}}\gamma )\,,
\end{array}\]
are the left-invariant one forms. The expressions for the vielbein coefficients are written by
\[\begin{array}{l}
E_0^0 = E_\tau ^0 = i\,L\,\Omega \,\cosh \rho \,,\\
E_1^1 = E_\rho ^1 = L\,\Omega \,,\\
E_2^2 = E_{{\phi _1}}^2 = L\,\Omega \,\sinh \rho \,,\\
E_3^3 = E_{{\phi _2}}^3 = L\,\Omega \,\sinh \rho \,\sin {\phi _1}\,,\\
E_4^4 = E_{{\phi _3}}^4 = L\,\Omega \,\sinh \rho \,\sin {\phi _1}\,\sin {\phi _2}\,,\\
E_5^5 = E_\theta ^5 = \sqrt {\frac{2}{3}} \,L\,\Omega \,,\\
E_6^6 = E_\alpha ^6 = \frac{1}{2}\,\sqrt {\frac{2}{3}} \,L\,\Omega \,\frac{{2\,\cos \theta \,\sin \beta }}{{\sqrt {3 - \cos 2\theta } }}\,,\\
E_8^8 = E_\gamma ^8 = \frac{1}{2}\,\sqrt {\frac{2}{3}} L\,\Omega \,\frac{{2\,\sin 2\theta \,\cos \alpha }}{{(3 - \cos 2\theta )}}\,,\\
E_9^9 = E_\phi ^9 = \frac{2}{3}\,L\,\Omega \,\left( {\frac{{1 - 3\,\cos 2\theta }}{{\cos 2\theta  - 3}}} \right)\,,\\
E_8^6 = E_\gamma ^6 =  - \frac{1}{2}\,\sqrt {\frac{2}{3}} \,L\,\Omega \,\frac{{2\,\cos \theta \,\cos \beta \,{\rm{sin}}\alpha }}{{\sqrt {3 - \cos 2\theta } }}\,,\\
E_6^7 = E_\alpha ^7 =  - \frac{1}{2}\,\sqrt {\frac{2}{3}} \,L\,\Omega \,\frac{{2\,\cos \theta \,\cos \beta }}{{\sqrt {3 - \cos 2\theta } }}\,,\\
E_8^7 = E_\gamma ^7 =  - \frac{1}{2}\,\sqrt {\frac{2}{3}} \,L\,\Omega \,\frac{{2\,\cos \theta \,\sin \beta \,{\rm{sin}}\alpha }}{{\sqrt {3 - \cos 2\theta } }}\,,\\
E_7^8 = E_\beta ^8 = \frac{1}{2}\,\sqrt {\frac{2}{3}} L\,\Omega \,\frac{{2\,\sin 2\theta }}{{(3 - \cos 2\theta )}}\,,\\
E_9^8 = E_\phi ^8 = \sqrt {\frac{2}{3}} L\,\Omega \,\frac{{2\,\sin 2\theta }}{{(3 - \cos 2\theta )}}\,,\\
E_7^9 = E_\beta ^9 =  - \frac{1}{3}\,L\,\Omega \,\left( {\frac{{1 - 3\,\cos 2\theta }}{{\cos 2\theta  - 3}}} \right)\,\frac{{4\,{{\cos }^2}\theta }}{{1 - 3\,\cos 2\theta }}\,,\\
E_8^9 = E_\gamma ^9 =  - \frac{1}{3}\,L\,\Omega \,\left( {\frac{{1 - 3\,\cos 2\theta }}{{\cos 2\theta  - 3}}} \right)\,\frac{{4\,{{\cos }^2}\theta \,\cos \alpha }}{{1 - 3\,\cos 2\theta }}\,,
\end{array}\]
where $\Omega$ is the warp factor from eq. (\ref{The warp factor at the IR point}), and $L$ is the AdS radius.
\section{Explicit form of the D7 $\kappa$ symmetry matrix}
\label{sec: Appendix-D}
The 8-form $\kappa$ symmetry matrix for the D7-brane has the following form
\begin{equation}
 {({\Gamma _\kappa })_{(8)}} = \frac{{\left( {{M_8}\,\sigma _3^4 + {M_6}\,\sigma _3^3 + {M_4}\,\sigma _3^2} \right)\,i\,{\sigma _2}}}{{\sqrt { - \det \left( {{\cal G} + {\cal B}} \right)} }}\,d\tau  \wedge d\rho  \wedge d{\phi _1} \wedge d{\phi _2} \wedge d{\phi _3} \wedge d\alpha  \wedge d\beta  \wedge d\gamma \,,
\end{equation}
where the Pauli matrices acts on the two component spinor $(\varepsilon_1^{\alpha},\varepsilon_2^{\beta})$, $\alpha,\,\beta=1,\dots,16$. The explicit form of the $M_4$ matrix is given by
\begin{align}
{M_4} &= {a_1}\,{\gamma _{\rho {\phi _2}{\phi _3}\beta }} + {a_2}\,{\gamma _{\rho {\phi _1}{\phi _3}\beta }} + {a_3}\,{\gamma _{\rho {\phi _1}{\phi _2}\beta }} + {a_4}\,{\gamma _{{\phi _1}{\phi _2}{\phi _3}\beta }} + {a_5}\,{\gamma _{\tau {\phi _2}{\phi _3}\beta }} + {a_6}\,{\gamma _{\tau {\phi _1}{\phi _3}\beta }}  \nonumber\\
 &  + {a_7}\,{\gamma _{\tau {\phi _1}{\phi _2}\beta }}+ {a_8}\,{\gamma _{\tau \rho {\phi _3}\beta }} + {a_9}\,{\gamma _{\tau \rho {\phi _2}\beta }} + {a_{10}}\,{\gamma _{\tau \rho {\phi _1}\beta }} + {a_{11}}\,{\gamma _{\rho {\phi _1}{\phi _2}{\phi _3}}} + {a_{12}}\,{\gamma _{\tau {\phi _1}{\phi _2}{\phi _3}}}\nonumber\\
 &+ {a_{13}}\,{\gamma _{\tau \rho {\phi _2}{\phi _3}}}+ {a_{14}}\,{\gamma _{\tau \rho {\phi _1}{\phi _3}}} + {a_{15}}\,{\gamma _{\tau \rho {\phi _1}{\phi _2}}}\,,
\end{align}
where $\gamma _{\rho {\phi _1}{\phi _2}\beta }=4! \,\gamma _{[\rho {\phi _1}{\phi _2}\beta ]}$ is an antisymmetric product of induced gamma matrices, and
\[
{a_0} = {B_{\gamma \phi }}\,{B_{\theta \alpha }} - {B_{\alpha \phi }}\,{B_{\theta \gamma }},
\]
\[
{a_1} =  - {a_0}\,\left( {{\partial _\tau }\phi \,{\partial _{{\phi _1}}}\theta  - {\partial _\tau }\theta \,{\partial _{{\phi _1}}}\phi } \right),\quad
{a_2} = {a_0}\,\left( {{\partial _\tau }\phi \,{\partial _{{\phi _2}}}\theta  - {\partial _\tau }\theta \,{\partial _{{\phi _2}}}\phi } \right),
\]
\[
{a_3} =  - {a_0}\,\left( {{\partial _\tau }\phi \,{\partial _{{\phi _3}}}\theta  - {\partial _\tau }\theta \,{\partial _{{\phi _3}}}\phi } \right),\quad
{a_{4}} =  - {a_0}\,\left( {{\partial _\rho }\phi \,{\partial _\tau }\theta  - {\partial _\rho }\theta \,{\partial _\tau }\phi } \right),\]
\[
{a_{5}} = {a_0}\,\left( {{\partial _\rho }\phi \,{\partial _{{\phi _1}}}\theta  - {\partial _\rho }\theta \,{\partial _{{\phi _1}}}\phi } \right),
\quad
{a_{6}} =  - {a_0}\,\left( {{\partial _\rho }\phi \,{\partial _{{\phi _2}}}\theta  - {\partial _\rho }\theta \,{\partial _{{\phi _2}}}\phi } \right),\]
\[
{a_{7}} = {a_0}\,\left( {{\partial _\rho }\phi \,{\partial _{{\phi _3}}}\theta  - {\partial _\rho }\theta \,{\partial _{{\phi _3}}}\phi } \right),\quad
{a_{8}} = {a_0}\,\left( {{\partial _{{\phi _1}}}\phi \,{\partial _{{\phi _2}}}\theta  - {\partial _{{\phi _1}}}\theta \,{\partial _{{\phi _2}}}\phi } \right),
\]
\[
{a_9} =  - {a_0}\,\left( {{\partial _{{\phi _1}}}\phi \,{\partial _{{\phi _3}}}\theta  - {\partial _{{\phi _1}}}\theta \,{\partial _{{\phi _3}}}\phi } \right),\quad
{a_{10}} = {a_0}\,\left( {{\partial _{{\phi _2}}}\phi \,{\partial _{{\phi _3}}}\theta  - {\partial _{{\phi _2}}}\theta \,{\partial _{{\phi _3}}}\phi } \right),
\]
\[
{a_{11}} = {B_{\beta \gamma }}\,\left( {{B_{\theta \alpha }}\,{\partial _\tau }\theta  - {B_{\alpha \phi }}\,{\partial _\tau }\phi } \right) + {B_{\alpha \beta }}\,\left( {{B_{\theta \gamma }}\,{\partial _\tau }\theta  - {B_{\gamma \phi }}\,{\partial _\tau }\phi } \right) - {a_0}\,\left( {{\partial _\beta }\phi \,{\partial _\tau }\theta  - {\partial _\beta }\theta \,{\partial _\tau }\phi } \right),
\]
\[
{a_{12}} = {B_{\beta \gamma }}\,\left( { - {B_{\theta \alpha }}\,{\partial _\rho }\theta  + {B_{\alpha \phi }}\,{\partial _\rho }\phi } \right) + {B_{\alpha \beta }}\,\left( { - {B_{\theta \gamma }}\,{\partial _\rho }\theta  + {B_{\gamma \phi }}\,{\partial _\rho }\phi } \right) + {a_0}\,\left( {{\partial _\beta }\phi \,{\partial _\rho }\theta  - {\partial _\beta }\theta \,{\partial _\rho }\phi } \right),
\]
\[
{a_{13}} = {B_{\beta \gamma }}\,\left( {{B_{\theta \alpha }}\,{\partial _{{\phi _1}}}\theta  - {B_{\alpha \phi }}\,{\partial _{{\phi _1}}}\phi } \right) + {B_{\alpha \beta }}\,\left( {{B_{\theta \gamma }}\,{\partial _{{\phi _1}}}\theta  - {B_{\gamma \phi }}\,{\partial _{{\phi _1}}}\phi } \right) - {a_0}\,\left( {{\partial _\beta }\phi \,{\partial _{{\phi _1}}}\theta  - {\partial _\beta }\theta \,{\partial _{{\phi _1}}}\phi } \right),
\]
\[
{a_{14}} = {B_{\beta \gamma }}\,\left( { - {B_{\theta \alpha }}\,{\partial _{{\phi _2}}}\theta  + {B_{\alpha \phi }}\,{\partial _{{\phi _2}}}\phi } \right) + {B_{\alpha \beta }}\,\left( { - {B_{\theta \gamma }}\,{\partial _{{\phi _2}}}\theta  + {B_{\gamma \phi }}\,{\partial _{{\phi _2}}}\phi } \right) + {a_0}\,\left( {{\partial _\beta }\phi \,{\partial _{{\phi _2}}}\theta  - {\partial _\beta }\theta \,{\partial _{{\phi _2}}}\phi } \right),
\]
\[
{a_{15}} = {B_{\beta \gamma }}\,\left( {{B_{\theta \alpha }}\,{\partial _{{\phi _3}}}\theta  - {B_{\alpha \phi }}\,{\partial _{{\phi _3}}}\phi } \right) + {B_{\alpha \beta }}\,\left( {{B_{\theta \gamma }}\,{\partial _{{\phi _3}}}\theta  - {B_{\gamma \phi }}\,{\partial _{{\phi _3}}}\phi } \right) - {a_0}\,\left( {{\partial _\beta }\phi \,{\partial _{{\phi _3}}}\theta  - {\partial _\beta }\theta \,{\partial _{{\phi _3}}}\phi } \right)\,.
\]
Here the $B_{\mu\nu}, \mu,\nu=0,\dots,7$, are the components of the second rank antisymmetric Kalb-Ramond B-field (\ref{NSKalbRamondBField}). The $M_6$ matrix is given by
\begin{align}
{M_6} &= {b_1}\,{\gamma _{\rho {\phi _1}{\phi _2}{\phi _3}\alpha \beta }} + {b_2}\,{\gamma _{\rho {\phi _1}{\phi _2}{\phi _3}\beta \gamma }} + {b_3}\,{\gamma _{\tau \rho {\phi _1}{\phi _2}\alpha \beta }} + {b_4}\,{\gamma _{\tau \rho {\phi _1}{\phi _2}\beta \gamma }} + {b_5}\,{\gamma _{\tau \rho {\phi _1}{\phi _2}{\phi _3}\alpha }} + {b_6}\,{\gamma _{\tau \rho {\phi _1}{\phi _2}{\phi _3}\beta }}\nonumber\\
 &+ {b_7}\,{\gamma _{\tau \rho {\phi _1}{\phi _2}{\phi _3}\gamma }} + {b_8}\,{\gamma _{\tau \rho {\phi _1}{\phi _3}\alpha \beta }} + {b_9}\,{\gamma _{\tau \rho {\phi _1}{\phi _3}\beta \gamma }} + {b_{10}}\,{\gamma _{\tau \rho {\phi _2}{\phi _3}\alpha \beta }} + {b_{11}}\,{\gamma _{\tau \rho {\phi _2}{\phi _3}\beta \gamma }} + {b_{12}}\,{\gamma _{\tau {\phi _1}{\phi _2}{\phi _3}\alpha \beta }}\nonumber\\
& + {b_{13}}\,{\gamma _{\tau {\phi _1}{\phi _2}{\phi _3}\beta \gamma }}\,,
\end{align}
where
\[
{b_1} = {B_{\theta \gamma }}\,{\partial _\tau }\theta  - {B_{\gamma \phi }}\,{\partial _\tau }\phi ,\quad
{b_2} = {B_{\theta \alpha }}\,{\partial _\tau }\theta  - {B_{\alpha \phi }}\,{\partial _\tau }\phi ,\]
\[
{b_3} = {B_{\theta \gamma }}\,{\partial _{{\phi _3}}}\theta  - {B_{\gamma \phi }}\,{\partial _{{\phi _3}}}\phi ,\quad
{b_4} = {B_{\theta \alpha }}\,{\partial _{{\phi _3}}}\theta  - {B_{\alpha \phi }}\,{\partial _{{\phi _3}}}\phi ,\]
\[
{b_5} = {B_{\beta \gamma }} + {B_{\theta \gamma }}\,{\partial _\beta }\theta  - {B_{\gamma \phi }}\,{\partial _\beta }\phi ,\quad
{b_6} =  - \left( {{B_{\alpha \gamma }} + {B_{\theta \gamma }}\,{\partial _\alpha }\theta  - {B_{\gamma \phi }}\,{\partial _\alpha }\phi  - {B_{\theta \alpha }}\,{\partial _\gamma }\theta  + {B_{\alpha \phi }}\,{\partial _\gamma }\phi } \right),
\]
\[
{b_7} = {B_{\alpha \beta }} - {B_{\theta \alpha }}\,{\partial _\beta }\theta  + {B_{\alpha \phi }}\,{\partial _\beta }\phi ,\quad
{b_8} =  - {B_{\theta \gamma }}\,{\partial _{{\phi _2}}}\theta  + {B_{\gamma \phi }}\,{\partial _{{\phi _2}}}\phi ,\quad
{b_9} =  - {B_{\theta \alpha }}\,{\partial _{{\phi _2}}}\theta  + {B_{\alpha \phi }}\,{\partial _{{\phi _2}}}\phi ,
\]
\[
{b_{10}} = {B_{\theta \gamma }}\,{\partial _{{\phi _1}}}\theta  - {B_{\gamma \phi }}\,{\partial _{{\phi _1}}}\phi ,\quad
{b_{11}} = {B_{\theta \alpha }}\,{\partial _{{\phi _1}}}\theta  - {B_{\alpha \phi }}\,{\partial _{{\phi _1}}}\phi ,\quad
{b_{12}} =  - {B_{\theta \gamma }}\,{\partial _\rho }\theta  + {B_{\gamma \phi }}\,{\partial _\rho }\phi ,\]
\[
{b_{13}} =  - {B_{\theta \alpha }}\,{\partial _\rho }\theta  + {B_{\alpha \phi }}\,{\partial _\rho }\phi \,.\]
The $M_8$ matrix takes the form
\begin{equation}
  {M_8} = {\gamma _{\tau \rho {\phi _1}{\phi _2}{\phi _3}\alpha \beta \gamma}}\,.
\end{equation}
The expressions for the induced gamma matrices are
\[\begin{array}{l}
{\gamma _\tau } = E_\tau ^0\,{\Gamma _0} + {\partial _\tau }\theta \,E_\theta ^5\,{\Gamma _5} + {\partial _\tau }\phi \,\left( {E_\phi ^8\,{\Gamma _8} + E_\phi ^9\,{\Gamma _9}} \right),\\
{\gamma _\rho } = E_\rho ^1\,{\Gamma _1} + {\partial _\rho }\theta \,E_\theta ^5\,{\Gamma _5} + {\partial _\rho }\phi \,\left( {E_\phi ^8\,{\Gamma _8} + E_\phi ^9\,{\Gamma _9}} \right),\\
{\gamma _{{\phi _1}}} = E_{{\phi _1}}^2\,{\Gamma _2} + {\partial _{\phi 1}}\theta \,E_\theta ^5\,{\Gamma _5} + {\partial _{{\phi _1}}}\phi \,\left( {E_\phi ^8\,{\Gamma _8} + E_\phi ^9\,{\Gamma _9}} \right),\\
{\gamma _{{\phi _2}}} = E_{{\phi _2}}^3\,{\Gamma _3} + {\partial _{{\phi _2}}}\theta \,E_\theta ^5\,{\Gamma _5} + {\partial _{{\phi _2}}}\phi \,\left( {E_\phi ^8\,{\Gamma _8} + E_\phi ^9\,{\Gamma _9}} \right),\\
{\gamma _{{\phi _3}}} = E_{{\phi _3}}^4\,{\Gamma _4} + {\partial _{{\phi _3}}}\theta \,E_\theta ^5\,{\Gamma _5} + {\partial _{{\phi _3}}}\phi \,\left( {E_\phi ^8\,{\Gamma _8} + E_\phi ^9\,{\Gamma _9}} \right),\\
{\gamma _\alpha } = E_\alpha ^6\,{\Gamma _6} + {\partial _\alpha }\theta \,E_\theta ^5\,{\Gamma _5} + {\partial _\alpha }\phi \,\left( {E_\phi ^8\,{\Gamma _8} + E_\phi ^9\,{\Gamma _9}} \right),\\
{\gamma _\beta } = E_\beta ^8\,{\Gamma _8} + E_\beta ^9\,{\Gamma _9} + {\partial _\beta }\theta \,E_\theta ^5\,{\Gamma _5} + {\partial _\beta }\phi \,\left( {E_\phi ^8\,{\Gamma _8} + E_\phi ^9\,{\Gamma _9}} \right),\\
{\gamma _\gamma } = E_\gamma ^7\,{\Gamma _7} + E_\gamma ^8\,{\Gamma _8} + E_\gamma ^9\,{\Gamma _9} + {\partial _\gamma }\theta \,E_\theta ^5\,{\Gamma _5} + {\partial _\gamma }\phi \,\left( {E_\phi ^8\,{\Gamma _8} + E_\phi ^9\,{\Gamma _9}} \right).
\end{array}\]
The kappa symmetry preserving condition (\ref{kappa symmetry preserving condition}) takes the form
\begin{equation}\label{D7 Kappa Symmetry System App}
\frac{1}{{\sqrt { - \det \left( {{\cal G} + {\cal B}} \right)} }}\,\left( {{M_8}\,\left( \begin{array}{l}
\,\,\,\,\varepsilon _2^\beta \\
 - \varepsilon _1^\alpha
\end{array} \right) + {M_6}\,\left( \begin{array}{l}
\varepsilon _2^\beta \\
\varepsilon _1^\alpha
\end{array} \right) + {M_4}\,\left( \begin{array}{l}
\,\,\,\,\varepsilon _2^\beta \\
 - \varepsilon _1^\alpha
\end{array} \right)} \right) = \left( \begin{array}{l}
\varepsilon _1^\alpha \\
\varepsilon _2^\beta
\end{array} \right)\,.
\end{equation}
This is an algebraic system for the components of the spinor. We can solve it by choosing a simpler D7-brane embedding ansatz, namely
\begin{equation}
\theta  = \theta (\rho ),\quad \phi  = \phi (\beta )\,.
\end{equation}
In this case the system (\ref{D7 Kappa Symmetry System App}) takes the form
\begin{equation}\label{D7 System1 App}
\sum\limits_{\beta  = 1}^{16} {{a_{\alpha \beta }}\,\epsilon _2^{(\beta )} = } \epsilon _1^{(\alpha )},\quad \alpha  = 1, \ldots ,16\,,
\end{equation}
\begin{equation}\label{D7 System2 App}
\sum\limits_{\beta  = 1}^{16} {{b_{\alpha \beta }}\,\epsilon _1^{(\beta )} = } \epsilon _2^{(\alpha )},\quad \alpha  = 1, \ldots ,16\,,
\end{equation}
where the coefficients $a_{\alpha\beta}$ and $b_{\alpha\beta}$ are functions of the coordinates. We can substitute the $\epsilon _2^{(\beta )}$ from (\ref{D7 System2 App}) into (\ref{D7 System1 App}). The resulting system is a homogenous system of 16 equations for the components $\epsilon _1^{(\alpha )}$:
\begin{equation}\label{D7KpappaSystemForTheHalfSpinors}
\sum\limits_{\beta  = 1}^{16} {{s_{\alpha \beta }}\,\varepsilon _1^{(\beta )} = } 0,\quad \alpha  = 1, \ldots ,16\,.
\end{equation}
We can write it explicitly as
\[\begin{array}{l}
{A_1}\,{\varepsilon _1} + {A_2}\,{\varepsilon _{11}} + {A_3}\,{\varepsilon _{16}} = 0,\quad {A_1}\,{\varepsilon _2} - {A_2}\,{\varepsilon _{12}} + {A_3}\,{\varepsilon _{15}} = 0\,,\\
{A_1}\,{\varepsilon _3} - {A_2}\,{\varepsilon _9} - {A_3}\,{\varepsilon _{14}} = 0,\quad \,\,{A_1}\,{\varepsilon _4} + {A_2}\,{\varepsilon _{10}} - {A_3}\,{\varepsilon _{13}} = 0\,,\\
{A_1}\,{\varepsilon _5} + {A_2}\,{\varepsilon _{15}} - {A_3}\,{\varepsilon _{12}} = 0,\quad {A_1}\,{\varepsilon _6} - {A_2}\,{\varepsilon _{16}} - {A_3}\,{\varepsilon _{11}} = 0\,,\\
{A_1}\,{\varepsilon _7} - {A_2}\,{\varepsilon _{13}} - {A_3}\,{\varepsilon _{10}} = 0,\quad {A_1}\,{\varepsilon _8} + {A_2}\,{\varepsilon _{14}} - {A_3}\,{\varepsilon _9} = 0\,,\\
{A_1}\,{\varepsilon _9} + {A_2}\,{\varepsilon _3} - {A_3}\,{\varepsilon _8} = 0,\quad \,\,\,\,\,{A_1}\,{\varepsilon _{10}} + {A_2}\,{\varepsilon _4} - {A_3}\,{\varepsilon _7} = 0\,,\\
{A_1}\,{\varepsilon _{11}} - {A_2}\,{\varepsilon _1} + {A_3}\,{\varepsilon _6} = 0\,,\quad \,{A_1}\,{\varepsilon _{12}} + {A_2}\,{\varepsilon _2} - {A_3}\,{\varepsilon _5} = 0\,,\\
{A_1}\,{\varepsilon _{13}} - {A_2}\,{\varepsilon _7} - {A_3}\,{\varepsilon _4} = 0\,,\quad \,{A_1}\,{\varepsilon _{14}} + {A_2}\,{\varepsilon _8} - {A_3}\,{\varepsilon _3} = 0\,,\\
{A_1}\,{\varepsilon _{15}} + {A_2}\,{\varepsilon _5} - {A_3}\,{\varepsilon _2} = 0\,,\quad \,{A_1}\,{\varepsilon _{16}} - {A_2}\,{\varepsilon _6} - {A_3}\,{\varepsilon _1} = 0\,,
\end{array}\]
where
\begin{align*}
{A_1} &=  - {D_0} - 37324800\,{({\partial _\rho }\theta )^2}{\cos ^2}\beta \,{\cos ^2}\theta \,{\left( {3 - \cos \left[ {2\theta } \right]} \right)^{7/2}}\,{\left( {1 - {\partial _\beta }\phi  + \left( {1 + 3\,{\partial _\beta }\phi } \right)\,\cos 2\theta } \right)^2}\,{\sin ^2}\alpha \\
 &- 82944\,{\left( {1 + 2\,{\partial _\beta }\phi } \right)^2}\,{({\partial _\rho }\theta )^2}\,{\cos ^4}\theta \,{\left( {3 - \cos 2\theta } \right)^{7/2}}\,{\sin ^2}\alpha \,{\sin ^2}\theta \\
 &- 223948800\,{\left( {1 + 2\,{\partial _\beta }\phi } \right)^2}\,{({\partial _\rho }\theta )^2}\,{\cos ^2}\beta \,{\cos ^4}\theta \,{\left( {3 - \cos 2\theta } \right)^{7/2}}\,{\sin ^2}\alpha \,{\sin ^2}\theta \\
 &- 335923200\,{\left( {1 + 2\,{\partial _\beta }\phi } \right)^2}\,{\cos ^4}\theta \,{\left( {3 - \cos 2\theta } \right)^{7/2}}\,{\sin ^2}\alpha \,{\sin ^2}\beta \,{\sin ^2}\theta \\
 &- 223948800\,{\left( {1 + 2\,{\partial _\beta }\phi } \right)^2}\,{({\partial _\rho }\theta )^2}\,{\cos ^4}\theta \,{\left( {3 - \cos 2\theta } \right)^{7/2}}\,{\sin ^2}\alpha \,{\sin ^2}\beta \,{\sin ^2}\theta \\
 &- 55987200\,{\partial _\beta }{\phi ^2}\,{\cos ^2}\alpha \,{\left( { - 56\,\sqrt 2 \,{\partial _\rho }\theta  + \cos \beta } \right)^2}\,{\left( {3 - \cos 2\theta } \right)^{9/2}}\,{\sin ^2}2\theta \\
 &- 18662400\,{\partial _\beta }{\phi ^2}\,{\cos ^2}\alpha \,{\left( {168 + \sqrt 2 \,{\partial _\rho }\theta \,\cos \beta } \right)^2}\,{\cos ^2}\theta \,{\left( {3 - \cos 2\theta } \right)^{5/2}}\,{\left( { - 7\,\sin \theta  + \sin 3\theta } \right)^2}\\
 &- 37324800\,{\partial _\beta }{\phi ^2}\,{({\partial _\rho }\theta )^2}\,{\cos ^2}\alpha \,{\cos ^2}\theta \,{\left( {3 - \cos 2\theta } \right)^{5/2}}\,{\sin ^2}\beta \,{\left( { - 7\,\sin \theta  + \sin 3\theta } \right)^2}\,,
\end{align*}
\[{A_2} =  - \sqrt {\frac{2}{3}} \,{\partial _\rho }\theta \,{A_3},\quad {D_0} = \frac{{729\,\sqrt 3 \,{{(3 - \cos 2\theta )}^{7/4}}}}{{4\,\sqrt[3]{2}\,{L^8}\,{{\sinh }^3}\rho \,\cosh \rho \,{{\sin }^2}{\phi _1}\,\sin {\phi _2}}}\,\sqrt { - \det ({\cal G} + {\cal B})} \,,\]
\[{A_3} = 64696320\,\sqrt 6 \,\left( {1 + 2\,{\partial _\beta }\phi } \right)\,{\cos ^3}\theta \,{\left( { - 3 + \cos 2\theta } \right)^4}\,\sin 2\alpha \,{\sin ^2}\theta \,{\partial _\beta }\phi \,{({\partial _\rho }\theta )^2}\,,\]
A homogenous system has a non-trivial solution when the determinant of its matrix is equal to zero. In our case we have to impose
\begin{equation}\label{D7DeterminantCondition}
  \det ({s_{\alpha \beta }}) = \frac{{{{\left( { - 3\,A_1^2 + A_3^2\,\left( {3 + 2\,{{({\partial _\rho }\theta )}^2}} \right)} \right)}^8}}}{{6561}} = 0\,.
\end{equation}
This allows us to find the following non-trivial solution of (\ref{D7KpappaSystemForTheHalfSpinors}):
\[\varepsilon _9^{(1)} = \frac{{ - \sqrt 2 \,{\partial _\rho }\theta \,\varepsilon _3^{(1)} + \sqrt 3 \,\varepsilon _8^{(1)}}}{{\sqrt {3 + 2\,{{({\partial _\rho }\theta )}^2}} }},\quad \varepsilon _{10}^{(1)} = \frac{{\sqrt 2 \,{\partial _\rho }\theta \,\varepsilon _4^{(1)} + \sqrt 3 \,\varepsilon _7^{(1)}}}{{\sqrt {3 + 2\,{{({\partial _\rho }\theta )}^2}} }}\,,\]
\[\varepsilon _{11}^{(1)} = \frac{{\sqrt 2 \,{\partial _\rho }\theta \,\varepsilon _1^{(1)} - \sqrt 3 \,\varepsilon _6^{(1)}}}{{\sqrt {3 + 2\,{{({\partial _\rho }\theta )}^2}} }},\quad \varepsilon _{12}^{(1)} = \frac{{ - \sqrt 2 \,{\partial _\rho }\theta \,\varepsilon _2^{(1)} - \sqrt 3 \,\varepsilon _5^{(1)}}}{{\sqrt {3 + 2\,{{({\partial _\rho }\theta )}^2}} }}\,,\]
\[\varepsilon _{13}^{(1)} = \frac{{\sqrt 3 \,\varepsilon _4^{(1)} - \sqrt 2 \,{\partial _\rho }\theta \,\varepsilon _7^{(1)}}}{{\sqrt {3 + 2\,{{({\partial _\rho }\theta )}^2}} }},\quad \varepsilon _{14}^{(1)} = \frac{{\sqrt 3 \,\varepsilon _3^{(1)} + \sqrt 2 \,{\partial _\rho }\theta \,\varepsilon _8^{(1)}}}{{\sqrt {3 + 2\,{{({\partial _\rho }\theta )}^2}} }}\,,\]
\[\varepsilon _{15}^{(1)} = \frac{{ - \sqrt 3 \,\varepsilon _2^{(1)} + \sqrt 2 \,{\partial _\rho }\theta \,\varepsilon _5^{(1)}}}{{\sqrt {3 + 2\,{{({\partial _\rho }\theta )}^2}} }},\quad \varepsilon _{16}^{(1)} = \frac{{ - \sqrt 3 \,\varepsilon _1^{(1)} - \sqrt 2 \,{\partial _\rho }\theta \,\varepsilon _6^{(1)}}}{{\sqrt {3 + 2\,{{({\partial _\rho }\theta )}^2}} }}\,.\]
Setting $\theta=const=0$ one finds
\begin{equation}
\begin{array}{l}
\varepsilon _9^{(1)} = \varepsilon _8^{(1)},\quad \varepsilon _{10}^{(1)} = \varepsilon _7^{(1)},\quad \varepsilon _{11}^{(1)} =  - \varepsilon _6^{(1)},\quad \varepsilon _{12}^{(1)} =  - \varepsilon _5^{(1)}\,,\\
\varepsilon _{13}^{(1)} =  - \varepsilon _7^{(1)},\quad \varepsilon _{14}^{(1)} = \varepsilon _8^{(1)},\quad \varepsilon _{15}^{(1)} = \varepsilon _5^{(1)},\quad \varepsilon _{16}^{(1)} =  - \varepsilon _6^{(1)}\,.
\end{array}
\end{equation}
There are 8 non-zero spinor solutions. This is what we expected -- the kappa symmetry removed exactly half of the spinor components. Setting $\theta=0$ in (\ref{D7DeterminantCondition}) also gives us $1 + {\partial _\beta }\phi  = 0$, which is solved by $\phi=-\beta+c$.
\section{Explicit form of the D5 $\kappa$ symmetry matrix}
\label{secC:Appendix-E}
The D5 6-form kappa symmetry matrix is given by
\begin{equation}\label{a}
{({\Gamma _\kappa })_{(6)}} = \frac{{{M_2}\,i\,\sigma _3^1\,{\sigma _2} + {M_4}\,i\,\sigma _3^2\,{\sigma _2}}}{{\sqrt { - \det \left( {{\cal G} + {\cal B}} \right)} }}\,d\tau  \wedge d\rho  \wedge d{\phi _2} \wedge d{\phi _3} \wedge d\beta  \wedge d\gamma \,,
\end{equation}
where we can write the matrices $M_2$ and $M_4$ as
\begin{equation}
{M_2} = \sum\limits_{i = 1}^{15} {{A_i}\,{a_i}\,} ,\;\;\;{\kern 1pt} {M_4} = \sum\limits_{i = 1}^{15} {{C_i}\,{c_i}} \,.
\end{equation}
The matrices $A_i$ are given by the following products of the induced gamma matrices:
\[
{A_1} = {\gamma _{\beta \gamma }},\quad {A_2} = {\gamma _{\rho \beta }},\quad {A_3} = {\gamma _{\rho \gamma }},\quad {A_4} = {\gamma _{\rho {\phi _2}}},\quad {A_5} = {\gamma _{\rho {\phi _3}}}\,,
\]
\[
{A_6} = {\gamma _{\tau \beta }},\quad {A_7} = {\gamma _{\tau \gamma }},\quad {A_8} = {\gamma _{\tau \rho }},\quad {A_9} = {\gamma _{\tau {\phi _2}}},\quad {A_{10}} = {\gamma _{\tau {\phi _3}}}\,,
\]
\[
{A_{11}} = {\gamma _{{\phi _2}\beta }},\quad {A_{12}} = {\gamma _{{\phi _2}\gamma }},\quad {A_{13}} = {\gamma _{{\phi _2}{\phi _3}}},\quad {A_{14}} = {\gamma _{{\phi _3}\beta }},\quad {A_{15}} = {\gamma _{{\phi _3}\gamma }}\,.
\]
The coefficients $a_i$ are expressed by the components of the induced $B$-field:
\[
{a_1} =  - {b_{\rho {\phi _3}}}\,{b_{\tau {\phi _2}}} + {b_{\rho {\phi _2}}}\,{b_{\tau {\phi _3}}} + {b_{\tau \rho }}\,{b_{{\phi _2}{\phi _3}}},\;\;\;{\kern 1pt} {a_2} =  - {b_{\tau {\phi _3}}}\,{b_{{\phi _2}\gamma }} + {b_{\tau \gamma }}\,{b_{{\phi _2}{\phi _3}}} + {b_{\tau {\phi _2}}}\,{b_{{\phi _3}\gamma }}\,,
\]
\[
{a_3} = {b_{\tau {\phi _3}}}\,{b_{{\phi _2}\beta }} - {b_{\tau \beta }}\,{b_{{\phi _2}{\phi _3}}} - {b_{\tau {\phi _2}}}\,{b_{{\phi _3}\beta }},\;\;\;{\kern 1pt} {a_4} = {b_{\beta \gamma }}\,{b_{\tau {\phi _3}}} + {b_{\tau \gamma }}\,{b_{{\phi _3}\beta }} - {b_{\tau \beta }}\,{b_{{\phi _3}\gamma }}\,,
\]
\[
{a_5} = {b_{\beta \gamma }}\,{b_{\tau {\phi _2}}} + {b_{\tau \gamma }}\,{b_{{\phi _2}\beta }} - {b_{\tau \beta }}\,{b_{{\phi _2}\gamma }},\;\;\;{\kern 1pt} {a_6} = {b_{\rho {\phi _3}}}\,{b_{{\phi _2}\gamma }} - {b_{\rho \gamma }}\,{b_{{\phi _2}{\phi _3}}} - {b_{\rho \phi }}\,{b_{{\phi _3}\gamma }}\,,
\]
\[
{a_7} =  - {b_{\rho {\phi _3}}}\,{b_{{\phi _2}\beta }} + {b_{\rho \beta }}\,{b_{{\phi _2}{\phi _3}}} + {b_{\rho {\phi _2}}}\,{b_{{\phi _3}\beta }},\;\;\;{\kern 1pt} {a_8} = {b_{\beta \gamma }}\,{b_{{\phi _2}{\phi _3}}} + {b_{{\phi _2}\gamma }}\,{b_{{\phi _3}\beta }} - {b_{{\phi _2}\beta }}\,{b_{{\phi _3}\gamma }}\,,
\]
\[
{a_9} = {b_{\beta \gamma }}\,{b_{\rho {\phi _3}}} + {b_{\rho \gamma }}\,{b_{{\phi _3}\beta }} - {b_{\rho \beta }}\,{b_{{\phi _3}\gamma }},\;\;\;{\kern 1pt} {a_{10}} = {b_{\beta \gamma }}\,{b_{\rho {\phi _2}}} + {b_{\rho \gamma }}\,{b_{{\phi _2}\beta }} - {b_{\rho \beta }}\,{b_{{\phi _2}\gamma }}\,,
\]
\[
{a_{12}} = {b_{\rho {\phi _3}}}\,{b_{\tau \beta }} - {b_{\rho \beta }}\,{b_{\tau {\phi _3}}} + {b_{\tau \rho }}\,{b_{{\phi _3}\beta }},\;\;\;{\kern 1pt} {a_{13}} =  - {b_{\rho \gamma }}\,{b_{\tau \beta }} + {b_{\rho \beta }}\,{b_{\tau \gamma }} + {b_{\beta \gamma }}\,{b_{\tau \rho }}\,,
\]
\[
{a_{14}} = {b_{\rho {\phi _2}}}\,{b_{\tau \gamma }} - {b_{\rho \gamma }}\,{b_{\tau {\phi _2}}} + {b_{\tau \rho }}\,{b_{{\phi _2}\gamma }},\;\;\;{\kern 1pt} {a_{15}} = {b_{\rho {\phi _2}}}\,{b_{\tau \beta }} - {b_{\rho \beta }}\,{b_{\tau {\phi _2}}} + {b_{\tau \rho }}\,{b_{{\phi _2}\beta }}\,.
\]
The expressions for the matrices $C_i$ are
\[
{C_1} = {\gamma _{\rho {\phi _2}\beta \gamma }},\quad {C_2} = {\gamma _{\rho {\phi _2}{\phi _3}\beta }},\quad {C_3} = {\gamma _{\rho {\phi _2}{\phi _3}\gamma }},\quad {C_4} = {\gamma _{\rho {\phi _3}\beta \gamma }},\quad {C_5} = {\gamma _{\tau \rho \beta \gamma }}\,,
\]
\[
{C_6} = {\gamma _{\tau \rho {\phi _2}\beta }},\quad {C_7} = {\gamma _{\tau \rho {\phi _2}\gamma }},\quad {C_8} = {\gamma _{\tau \rho {\phi _2}{\phi _3}}},\quad {C_9} = {\gamma _{\tau \rho {\phi _3}\beta }},\quad {C_{10}} = {\gamma _{\tau \rho {\phi _3}\gamma }}\,,
\]
\[
{C_{11}} = {\gamma _{\tau {\phi _2}\beta \gamma }},\quad {C_{12}} = {\gamma _{\tau {\phi _2}{\phi _3}\beta }},\quad {C_{13}} = {\gamma _{\tau {\phi _2}{\phi _3}\gamma }},\quad {C_{14}} = {\gamma _{\tau {\phi _3}\beta \gamma }},\quad {C_{15}} = {\gamma _{{\phi _2}{\phi _3}\beta \gamma }}\,.
\]
The expressions for the coefficients $c_i$ are given by
\[
{c_1} = {b_{\tau {\phi _3}}},\quad{c_2} = {b_{\tau \gamma }},\quad {c_3} =  - {b_{\tau \beta }},\quad {c_4} =  - {b_{\tau {\phi _2}}},\quad {c_5} = {b_{{\phi _2}{\phi _3}}},\]
\[{c_6} =  - {b_{{\phi _3}\gamma }},\quad {c_7} = {b_{{\phi _3}\beta }}\,,
\quad {c_8} = {b_{\beta \gamma }}\,,\quad {c_9} = {b_{{\phi _2}\gamma }},\quad {c_{10}} =  - {b_{{\phi _2}\beta }},\]
\[\quad {c_{11}} =  - {b_{\rho {\phi _3}}},\quad {c_{12}} =  - {b_{\rho \gamma }},\quad {c_{13}} = {b_{\rho \beta }}\,,
\quad
{c_{14}} = {b_{\rho {\phi _2}}},\quad {c_{15}} = {b_{\tau \rho }}\,.
\]
We also have the components of the $B_{(2)}$-form pullback,
\begin{equation}
\mathcal{B}_{(2)}=P[{B_{(2)}}] = \sum\limits_{a,\,b = 0}^5 {{b_{ab}}\,da \wedge db} \,,
\end{equation}
given by
\[
{b_{\tau {\phi _3}}} = \left( {{B_{\theta \alpha }}\,{\partial _\tau }\theta  - {B_{\alpha \phi }}\,{\partial _\tau }\phi } \right)\,{\partial _{{\phi _3}}}\alpha  - {B_{\theta \alpha }}\,{\partial _\tau }\alpha \,{\partial _{{\phi _3}}}\theta  + {B_{\alpha \phi }}\,{\partial _\tau }\alpha \,{\partial _{{\phi _3}}}\phi \,,
\]
\[
{b_{\tau {\phi _2}}} = \left( {{B_{\theta \alpha }}\,{\partial _\tau }\theta  - {B_{\alpha \phi }}\,{\partial _\tau }\phi } \right)\,{\partial _{{\phi _2}}}\alpha  - {B_{\theta \alpha }}\,{\partial _\tau }\alpha \,{\partial _{{\phi _2}}}\theta  + {B_{\alpha \phi }}\,{\partial _\tau }\alpha \,{\partial _{{\phi _2}}}\phi \,,
\]
\[
{b_{\tau \rho }} =  - {B_{\theta \alpha }}\,{\partial _\rho }\theta \,{\partial _\tau }\alpha  + {B_{\alpha \phi }}\,{\partial _\rho }\phi \,{\partial _\tau }\alpha  + {\partial _\rho }\alpha \,({B_{\theta \alpha }}\,{\partial _\tau }\theta  - {B_{\alpha \phi }}\,{\partial _\tau }\phi )\,,
\]
\[
{b_{\rho {\phi _2}}} = \left( {{B_{\theta \alpha }}\,{\partial _\rho }\theta  - {B_{\alpha \phi }}\,{\partial _\rho }\phi } \right)\,{\partial _{{\phi _2}}}\alpha  - {B_{\theta \alpha }}\,{\partial _\rho }\alpha \,{\partial _{{\phi _2}}}\theta  + {B_{\alpha \phi }}\,{\partial _\rho }\alpha \,{\partial _{{\phi _2}}}\phi \,,
\]
\[
{b_{\rho {\phi _3}}} = \left( {{B_{\theta \alpha }}\,{\partial _\rho }\theta  - {B_{\alpha \phi }}\,{\partial _\rho }\phi } \right){\partial _{{\phi _3}}}\alpha  - {B_{\theta \alpha }}\,{\partial _\rho }\alpha \,{\partial _{{\phi _3}}}\theta  + {B_{\alpha \phi }}\,{\partial _\rho }\alpha \,{\partial _{{\phi _3}}}\phi \,,
\]
\[
{b_{{\phi _2}{\phi _3}}} = \left( {{B_{\theta \alpha }}\,{\partial _{{\phi _2}}}\theta  - {B_{\alpha \phi }}\,{\partial _{{\phi _2}}}\phi } \right)\,{\partial _{{\phi _3}}}\alpha  - {B_{\theta \alpha }}\,{\partial _{{\phi _2}}}\alpha \,{\partial _{{\phi _3}}}\theta  + {B_{\alpha \phi }}\,{\partial _{{\phi _2}}}\alpha \,{\partial _{{\phi _3}}}\phi \,,
\]
\[
{b_{\tau \beta }} = \left( {{B_{\alpha \beta }} - {B_{\theta \alpha }}\,{\partial _\beta }\theta  + {B_{\alpha \phi }}\,{\partial _\beta }\phi } \right)\,{\partial _\tau }\alpha  + {\partial _\beta }\alpha \,\left( {{B_{\theta \alpha }}\,{\partial _\tau }\theta  - {B_{\alpha \phi }}\,{\partial _\tau }\phi } \right)\,,
\]
\[
{b_{\rho \beta }} = \left( {{B_{\alpha \beta }} - {B_{\theta \alpha }}\,{\partial _\beta }\theta  + {B_{\alpha \phi }}\,{\partial _\beta }\phi } \right)\,{\partial _\rho }\alpha  + {\partial _\beta }\alpha \,\left( {{B_{\theta \alpha }}\,{\partial _\rho }\theta  - {B_{\alpha \phi }}\,{\partial _\rho }\phi } \right)\,,
\]
\[
{b_{{\phi _3}\beta }} = \left( {{B_{\alpha \beta }} - {B_{\theta \alpha }}\,{\partial _\beta }\theta  + {B_{\alpha \phi }}\,{\partial _\beta }\phi } \right)\,{\partial _{{\phi _3}}}\alpha  + {\partial _\beta }\alpha \,\left( {{B_{\theta \alpha }}\,{\partial _{{\phi _3}}}\theta  - {B_{\alpha \phi }}\,{\partial _{{\phi _3}}}\phi } \right)\,,
\]
\[
{b_{{\phi _2}\beta }} = \left( {{B_{\alpha \beta }} - {B_{\theta \alpha }}\,{\partial _\beta }\theta  + {B_{\alpha \phi }}\,{\partial _\beta }\phi } \right)\,{\partial _{{\phi _2}}}\alpha  + {\partial _\beta }\alpha \,\left( {{B_{\theta \alpha }}\,{\partial _{{\phi _2}}}\theta  - {B_{\alpha \phi }}\,{\partial _{{\phi _2}}}\phi } \right)\,,
\]
\[
{b_{\rho \gamma }} = \left( {{B_{\alpha \gamma }} - {B_{\theta \alpha }}\,{\partial _\gamma }\theta  + {B_{\alpha \phi }}\,{\partial _\gamma }\phi } \right)\,{\partial _\rho }\alpha  + \left( {{B_{\theta \gamma }} + {B_{\theta \alpha }}{\partial _\gamma }\alpha } \right)\,{\partial _\rho }\theta  - \left( {{B_{\gamma \phi }} + {B_{\alpha \phi }}{\partial _\gamma }\alpha } \right)\,{\partial _\rho }\phi \,,
\]
\[
{b_{\tau \gamma }} = \left( {{B_{\alpha \gamma }} - {B_{\theta \alpha }}\,{\partial _\gamma }\theta  + {B_{\alpha \phi }}\,{\partial _\gamma }\phi } \right)\,{\partial _\tau }\alpha  + \left( {{B_{\theta \gamma }} + {B_{\theta \alpha }}\,{\partial _\gamma }\alpha } \right)\,{\partial _\tau }\theta  - \left( {{B_{\gamma \phi }} + {B_{\alpha \phi }}\,{\partial _\gamma }\alpha } \right)\,{\partial _\tau }\phi \,,
\]
\[
{b_{{\phi _2}\gamma }} = \left( {{B_{\alpha \gamma }} - {B_{\theta \alpha }}\,{\partial _\gamma }\theta  + {B_{\alpha \phi }}\,{\partial _\gamma }\phi } \right)\,{\partial _{{\phi _2}}}\alpha  + \left( {{B_{\theta \gamma }} + {B_{\theta \alpha }}\,{\partial _\gamma }\alpha } \right)\,{\partial _{{\phi _2}}}\theta  - \left( {{B_{\gamma \phi }} + {B_{\alpha \phi }}\,{\partial _\gamma }\alpha } \right)\,{\partial _{{\phi _2}}}\phi \,,
\]
\[
{b_{{\phi _3}\gamma }} = \left( {{B_{\alpha \gamma }} - {B_{\theta \alpha }}\,{\partial _\gamma }\theta  + {B_{\alpha \phi }}\,{\partial _\gamma }\phi } \right)\,{\partial _{{\phi _3}}}\alpha  + \left( {{B_{\theta \gamma }} + {B_{\theta \alpha }}\,{\partial _\gamma }\alpha } \right)\,{\partial _{{\phi _3}}}\theta  - \left( {{B_{\gamma \phi }} + {B_{\alpha \phi }}\,{\partial _\gamma }\alpha } \right)\,{\partial _{{\phi _3}}}\phi \,,
\]
\[
{b_{\beta \gamma }} =  {B_{\theta \gamma }}\,{\partial _\beta }\theta - {B_{\beta \gamma }} - {B_{\gamma \phi }}\,{\partial _\beta }\phi  - \left( {{B_{\alpha \beta }} - {B_{\theta \alpha }}\,{\partial _\beta }\theta  + {B_{\alpha \phi }}\,{\partial _\beta }\phi } \right)\,{\partial _\gamma }\alpha +{\partial _\beta }\alpha \,\left( {{B_{\alpha \gamma }} - {B_{\theta \alpha }}\,{\partial _\gamma }\theta  + {B_{\alpha \phi }}\,{\partial _\gamma }\phi } \right)\,,
\]
where the explicit form of the components of the $B$-field are given in eq. (\ref{NSKalbRamondBField}).
%

\end{document}